\documentclass[a4paper,11pt]{article}
\usepackage{jheppub}
\usepackage{xspace}			   
\usepackage[english]{babel}
\usepackage[T1]{fontenc}
\usepackage{lmodern}

\usepackage{setspace}
\usepackage{ragged2e}
\usepackage{subfig}
\usepackage[all]{xy}

\usepackage{pdfpages}
\usepackage{amsmath, mathtools}
\usepackage{amssymb,amsfonts,dsfont}
\usepackage{mathrsfs}
\usepackage{amsthm}
\usepackage{bm}				
\usepackage{bbm,upgreek}
\usepackage{makecell}
\usepackage{cellspace}
\usepackage[sort&compress,numbers]{natbib}
\usepackage{url}
\usepackage{doi}

\newcommand{\bra}[1]{\langle #1 |} 
\newcommand{\ket}[1]{| #1 \rangle } 
\definecolor{cbl}{rgb}{0,0,1}
 
\definecolor{crd}{rgb}{1,0,0}
 
\newcommand{\upd}{\mathrm{d}}
\newcommand{\tr}{\mathrm{tr}}

\newcommand{\ie}[0]{\textit{i.e. }}
\newcommand{\eg}[0]{\textit{e.g. }}

\newcommand\e{\mathrm{e}}

\newcommand\hpi{\hat{\pi}}
\newcommand\hphi{\hat{\phi}}

\newcommand{\normord}[1]{:\mathrel{#1}:}

\definecolor{nblue}{rgb}{0.06,0.3,0.73}
\definecolor{nblack}{rgb}{0,0,0}
\definecolor{nred}{rgb}{0.9,0.1,0.1}

\DeclareMathOperator*{\argmin}{argmin}

\newcommand{\id}{{\mathds 1}}

\newcommand{\D}{\Delta}

\def\bea{\begin{eqnarray}} \def\eea{\end{eqnarray}}
\def\be{\begin{equation}} \def\ee{\end{equation}} 
\def\nn{\nonumber}
\newcommand{\order}{{\mathcal O}}
\newcommand{\Dmu}{{\mathcal D}_L}
\newcommand{\braket}[1]{\left\langle #1 \right\rangle } 
\definecolor{cbl}{rgb}{0,0,1}
\definecolor{crd}{rgb}{1,0,0}

\newcommand{\cross}[2]{%
    \draw[blue, thick] (#1-0.1,#2-0.1) -- (#1+0.1,#2+0.1);
    \draw[blue, thick] (#1-0.1,#2+0.1) -- (#1+0.1,#2-0.1);
}
\usepackage{tikz}
\tikzset{
	vtx/.style={
		circle,
		draw=blue,
		fill=blue,
		inner sep=1pt
	},
	wcirc/.style={
		circle,
		draw=white,
		fill=white,
		inner sep=2pt
	},
	bcirc/.style={
		circle,
		draw=black,
		fill=black,
		inner sep=1pt
	},
	dcirc/.style={
		circle,
		draw=blue,
		fill=blue,
		inner sep=1pt
	},
	rcirc/.style={
		circle,
		draw=black,
		fill=black,
		inner sep=1pt
	},
	phi/.style={
		thick
	},
	sigma/.style={
		thick,
		[dashed, blue]
	},
	vl1/.style={
		thick,
		blue
	},
	vl2/.style={
		thick,
		dashed,
		blue
	},
	valign/.style={
		baseline={([yshift=-.55ex]current bounding box.center)}
	}
}
\usetikzlibrary{decorations.pathmorphing}

\theoremstyle{remark} %

\title{A relativistic continuous matrix product state study of field theories with defects}

\author{Karan Tiwana,}
\emailAdd{karantiwana21693@gmail.com}
\author{Edoardo Lauria,}
\emailAdd{edoardo.lauria@minesparis.psl.eu}
\author{Antoine Tilloy}
\emailAdd{antoine.tilloy@minesparis.psl.eu}

\affiliation{Laboratoire de Physique de l’École Normale Supérieure, Mines Paris - PSL, Inria, CNRS, ENS-PSL, Sorbonne Université, PSL Research University, Paris, France}

\abstract{
We propose a method to compute expectation values in 1+1-dimensional massive Quantum Field Theories (QFTs) with line defects using Relativistic Continuous Matrix Product State (RCMPS). Exploiting Euclidean invariance, we use a quantization scheme where (imaginary) time runs perpendicularly to the defect. With this choice, correlation functions of local operators in the presence of the defect can be computed as expectation values of extended operators in the no-defect vacuum, which can be approximated by a homogeneous RCMPS. We demonstrate the effectiveness of this machinery by computing correlation functions of local \emph{bulk} and \emph{defect} operators in $\phi^4$ theory with a magnetic line defect, in perturbative, strong coupling, critical, and symmetry-broken regimes.
}

\begin{document}

\maketitle

\section{Introduction}
Relativistic scalar quantum field theories (QFT) in $1+1$ space-time dimensions are an excellent sandbox for theorists: they can be defined rigorously in a fairly straightforward way, and yet are generically difficult to solve non-perturbatively. Intuitively, such theories are numerically challenging because their number of degrees of freedom grows both when the short distance (UV) cutoff and long distance (IR) cutoff are lifted. Nonetheless, a tremendous amount of progress has been made in recent years to compute local expectation values of low dimensional QFT beyond perturbation theory. In this article, our objective is to go beyond simple local expectation values, and get numerically precise predictions in 1+1 dimensional QFTs with \emph{defects}. This has mainly two interests: 1) we get numerical access to new non-trivial physics, like that of impurities, and 2) we can benchmark non-perturbative methods on a more difficult example.

\subsection{Defects}
In the general case, \ie in a $d$-dimensional space-time, a $p$-dimensional defect is a modification of the $d$-dimensional action (the ``bulk theory'') that has support on $p$-dimensional submanifold $B$ of $\mathbb{R}^d$. More explicitly, but at a (so far) non-rigorous level, we consider ``bulk'' QFTs with an action of the form
\begin{equation}
    S_\text{bulk}(\phi) = \int_{\mathbb{R}^d} \frac{(\nabla \phi)^2}{2} + V(\phi)\,,
\end{equation}
where we will later fix $d=2$. The ``defect'' QFT is specified by the action
\begin{equation}
    S(\phi) = S_\text{bulk}(\phi) + S_\text{defect}(\phi) = S_\text{bulk}(\phi) + \int_B f(\phi)\,.
\end{equation}
Expectation values can be computed with this modified action in the same way as for a standard QFT. Again non-rigorously, in Euclidean signature we have that:
\begin{equation}
    \langle \phi(x_1)\cdots \phi(x_n)\rangle_\text{defect} = \frac{\int \mathcal{D}\phi \; \phi(x_1)\cdots \phi(x_n) \; \e^{-S(\phi)} }{\int \mathcal{D}\phi  \; \e^{-S(\phi)} }= \frac{\langle \phi(x_1)\cdots \phi(x_n) \, \e^{-\int_B f(\phi)}\,\rangle_\text{bulk}}{\langle \e^{-\int_B f(\phi)}\,\rangle_\text{bulk}}\,.
\end{equation}
Physically, defects can be used to model boundaries, interfaces, and impurities which are very common in realistic low-energy systems (\eg in the Kondo effect \cite{Wilson:1974mb,KONDO1970183} or surface criticality, see \cite{Cardy1996,DiFrancesco:1997nk} for review). Considering defects is also useful in the formal study of QFT, and \eg allows for a characterization of symmetries that are not captured by local QFT operators \cite{Gaiotto:2014kfa} (see \cite{Cordova:2022ruw} for a recent review).

There has been a lot of progress in understanding the role of (non-topological) defects in QFT, mostly when the bulk action QFT is conformally invariant, both in $d=2$ \cite{Lewellen:1991tb,Runkel:1998he,Cardy:2004hm} and higher (see e.g.~\cite{Liendo:2012hy,Billo:2016cpy} for early studies) with a recent explosion \cite{Lauria:2020emq,Cuomo:2021kfm,Nishioka:2022qmj,Gimenez-Grau:2022ebb,Giombi:2022vnz,Rodriguez-Gomez:2022gif,Bianchi:2022sbz,Aharony:2022ntz,Cuomo:2022xgw,Aharony:2023amq,Barrat:2023ivo,Bianchi:2023gkk,Nagar:2024mjz,Diatlyk:2024zkk,Kravchuk:2024qoh,Cuomo:2024psk}.
However, much less is known for gapped bulk actions (conformally invariant at short but not large distance) beyond perturbative regimes. 

In this work, we study a class of (non-topological) line defects for the massive $\phi^4$ theory: the so called magnetic line defects. As reviewed in \cite{PhysRevB.61.15152,PhysRevB.68.064419,PhysRevLett.96.036601,PhysRevX.3.031010,PhysRevB.95.014401,2014arXiv1412.3449A} and discussed more recently in \cite{Cuomo:2021kfm, Cuomo:2022xgw}, such defects play an important role in condensed matter systems where they capture the physics of magnetic-like impurities. For conformal QFTs in $d\geq 2$, magnetic line defects can be studied with a variety of techniques, starting from perturbation theory and conformal bootstrap \cite{Nishioka:2022qmj,Gimenez-Grau:2022ebb,Giombi:2022vnz,Rodriguez-Gomez:2022gif,Bianchi:2022sbz,Franchi:2022rkx,Rodriguez-Gomez:2022gbz,Popov:2022nfq,Gimenez-Grau:2022czc,Pannell:2023pwz,Hu:2023ghk,Zhou:2023fqu,Dey:2024ilw} and more recently fuzzy sphere regularization \cite{Cuomo:2024psk}.

\subsection{The variational method}

Independently of defects, the two historical methods to solve generic QFTs are perturbation theory and lattice Monte Carlo, both of which have limitations. In the first case, one can lift both UV and IR cutoffs, but results are given as a divergent power series in the coupling (which can be Borel resummed in favorable cases, see e.g.~\cite{Serone:2018gjo,Serone:2019szm}). In the second case, results are fully non-perturbative, at the cost of hard UV and IR cutoffs (a finite size space-time lattice), and statistical errors. There is now a plethora of less traditional methods, including the modern S-matrix bootstrap (see \cite{Kruczenski:2022lot} for a recent review and references), functional renormalization \cite{rosten2012frgreview,balog2019FRGconvergence}, and possibly soon even direct quantum simulation \cite{preskill2018simulatingquantumfieldtheory,brennen2015qsimulationofQFT,hardy2024optimizedquantumsimulationalgorithms}, that could in principle be modified to account for defects.

The variational approach yields another class of non-perturbative methods (including Hamiltonian truncation \cite{Hogervorst:2014rta,rychkov2015,Elias-Miro:2017xxf,elias-miro2017NLOrenormalization}) which we now focus on. The idea is fairly 
simple: 
treat a QFT in canonical quantization simply as a particular quantum mechanical system, with a Hilbert space $\mathscr{H}$ and Hamiltonian
\begin{equation}
    H_\text{bulk} = \int_\mathbb{R} h := \int_\mathbb{R}  \frac{ \hpi^2}{2} + \frac{ (\nabla \hphi)^2}{2} + V(\hphi)~, \qquad \text{with} \quad [\hphi(x),\hpi(y)] = i \delta(x-y)\,,
\end{equation}
and approximate its ground state (the non-perturbative vacuum) $\ket{\Omega}$ by minimizing the energy density over a manageable submanifold $\mathcal{M}$ of states:
\begin{equation}
    \ket{\Omega} \simeq \ket{\psi_0} := \argmin_{\ket{\psi} \in \mathcal{M}\subset \mathscr{H}} \frac{\bra{\psi} h \ket{\psi}}{\bra{\psi} \psi \rangle}\,.
\end{equation}
The output of a \emph{bona fide} variational approach is an approximate ground state $\ket{\psi_0}$ and an efficient algorithm to evaluate expectation values $\bra{\psi_0} \hphi(x_1) \cdots \hphi(x_n)\ket{\psi_0}$ of interest. This strategy is particularly adapted to $1+1$ dimensional scalar QFT because most models have a finite energy density in the vacuum, and thus the optimization problem is \emph{in principle} well posed. 

A particularly powerful class of states to use in the variational approach are relativistic continuous matrix product states (RCMPS) \cite{Tilloy:2021yre,Tilloy:2021hhb}. They have a number of appealing features: 1) they are defined directly in the continuum (without UV cutoff), 2) their thermodynamic limit can be taken easily (no IR cutoff), 3) their expressiveness (the dimension of $\mathcal{M}$) can be arbitrarily increased, 4) local expectation values can be computed (numerically) exactly, and 5) optimizing over them is practically efficient. These states have been successfully used to study vacuum expectation values in the $\phi^4$ \cite{Tilloy:2021hhb}, Sine-Gordon, and Sinh-Gordon models \cite{Tilloy:2022kcn}.

\subsection{Rotating the defect, from impurity to extended operator}\label{sec:Wick_Rotation}
In $d=1+1$ space-time dimension, we can consider line defects (\ie $p=1$): a one dimensional deformation of the two dimensional action, which we take to be a straight line. In the Hamiltonian picture, such a defect is naturally interpreted as an impurity, namely a perturbation of the Hamiltonian localized at one point $x_0$ in space, and persistent in time:
\begin{equation}
    H = H_\text{bulk} + H_\text{defect} = H_\text{bulk} + f\left[\hphi(x_0)\right]\,.
\end{equation}
If the defect is only a finite or semi-infinite line, this corresponds to a \emph{local quench}, that is an impurity that is turned on only for some time. 

For generic quantum many-body systems, such impurities/quenches give rise to rich physics but introduce challenges with the variational method. First, translation invariance in space is lost. Second, in the local quench case, the quantum state needed to evaluate expectation values becomes time dependent, and thus needs to be evolved \eg with the time dependent variational principle (TDVP) \cite{haegeman2011_original_tdvpMPS,vanderstraeten2019_tangentspace}.

\begin{figure}[htp]
    \centering
    \begin{tikzpicture}[scale=1.5]

    \tikzset{line width=0.8pt}
    \definecolor{bulkblue}{rgb}{0.1, 0.1, 0.7}
    
    \begin{scope}
      \node at (-0.5,1.5) {\footnotesize a)};
      
      \draw[->] (0,0) -- (0,1.7) node[left] {$\tau$};
      \draw[->] (0,0) -- (2.3,0) node[below] {$x$};
      
      \draw[bulkblue, thick] (0.5,-0.1) -- (0.5,1.7);
      \node[bulkblue, left] at (0.5,0.3) {$B$};
      \node[bulkblue] at (1.1,1.5) {$e^{-\int_B f(\phi)}$};
      
      \node[right] at (1.2,0.5) {$e^{-S_{\text{bulk}}}$};
    \end{scope}

    \begin{scope}[xshift=3cm]
      \node at (-0.5,1.5) {\footnotesize b)};
      
      \draw[->] (0,0) -- (0,1.7) node[left] {$\tau$};
      \draw[->] (0,0) -- (2.3,0) node[below] {$x$};
      
      \draw[bulkblue, thick] (-0.1,0.5) -- (2.3,0.5);
      \node[bulkblue, below] at (0.3,0.5) {$B$};
      
      \node[bulkblue] at (1.9,0.8) {$e^{-\int_B f(\phi)}$};
      
      \node[right] at (0.5,1.5) {$e^{-S_{\text{bulk}}}$};
    \end{scope}
    
    \begin{scope}[xshift=6.5cm]
      \node at (-0.5,1.5) {\footnotesize c)};
      
      \draw[->] (0,0) -- (0,1.7) node[left] {$\tau$};
      \draw[->] (0,0) -- (2.3,0) node[below] {$x$};
      
      \draw[bulkblue, thick] (-0.3,0.8) -- (2.3,0.8);
      \node[bulkblue] at (-0.4,0.68) {$B$};
      
    
      \draw[->] (0.7,1.6) -- (0.7,0.86) node[midway, right] {$e^{-H\tau}$};
      \draw[->] (0.7,0.1) -- (0.7,0.74) node[midway, right] {$e^{-H\tau}$};
      
      \draw[thick] (-0.1,0.85) -- (2.2,0.85);
      \node[below] at (2.0,0.80) {$\bra{\Omega}$};
      
      \draw[thick] (-0.1,0.75) -- (2.2,0.75);
      \node[above] at (2.2,0.86) {$\ket{\Omega}$};
    \end{scope}
    \end{tikzpicture}
    \caption{a) original defect, interpreted as an impurity, b) equivalent rotated defect, interpreted as an extended operator c) representation of the defect expectation value in the Hamiltonian formalism: the bulk dynamics from $+\infty$ and $-\infty$  is equivalent to a projection of the state at fixed $\tau$ to the bulk ground state $\ket{\Omega}$.}
    \label{fig:defect_rotation}
\end{figure}
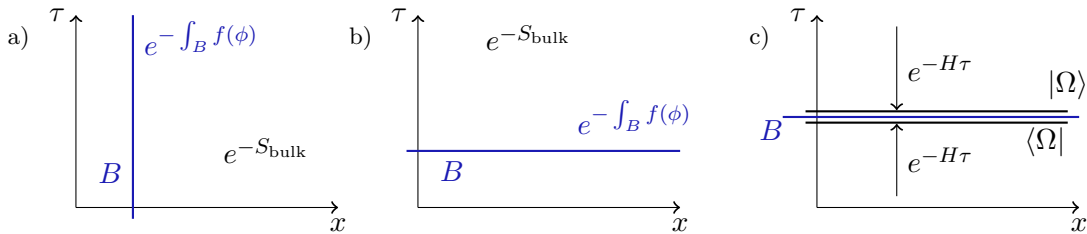

Fortunately, a simple observation makes a plain variational study possible for large subset of defects and operator insertions. For a relativistic QFT in imaginary time, \ie a Euclidean invariant theory, we can rotate the quantization direction, exchange the role of space and time, and make the defect spacelike instead of timelike (see Fig. \ref{fig:defect_rotation}). In this new picture, time evolution is generated by the bulk Hamiltonian, and expectation values are simply evaluated by applying an extended operator to the translation-invariant bulk vacuum, $\ket{\Omega}$:
\begin{equation}\label{eq:rotated_defect_expectations}
   \langle \phi(x_1) \cdots \phi(x_n)\rangle_\text{defect} = \frac{\bra{\Omega} \, \hphi(x_1) \cdots \hphi(x_n)\,  \e^{-\int_B f(\hphi)} \ket{\Omega}}{\bra{\Omega} \e^{-\int_B f(\hphi)} \ket{\Omega}}\,.
\end{equation}

This representation is convenient because $\ket{\Omega}$ can be replaced by its variational approximation as a RCMPS $\ket{\psi_0}$, simply computed from the bulk theory. The last ingredient is technical, non-trivial, and specific to RCMPS: if the operator insertions and defect are aligned (same imaginary time), and if $f$ is \emph{linear} (\ie a so called magnetic defect), the right-hand side of \eqref{eq:rotated_defect_expectations} is efficiently computable at the same asymptotic cost with the defect than in the vanilla no defect case. All the complex optimization toolbox required to find RCMPS ground states for translation invariant models can thus be reused without modification. All that is needed is a particular post-processing of this ground state, which is comparatively cheap.

\subsection{Advantages and limitations}
 Although we test it only for a $\phi^4$ potential here, the method can be used for any bulk model with a potential $V(\phi)$ for which RCMPS optimization algorithms have been developed. Currently, this includes all polynomial potentials, real exponentials $e^{\beta \phi}$, and imaginary exponentials $e^{i\beta \phi}$ with $\beta < \sqrt{4\pi}$. 
 
 The method we present has the same cost and precision as standard RCMPS. At short distances, in the deep UV, RCMPS behave like the free Fock vacuum. For the super-renormalizable bulk theories we consider, this is also the behavior of the true ground state, and thus our method has excellent precision in this limit. As we will see, it compares favorably with third-order perturbation theory. At large distance, RCMPS provide good approximations of gapped theories even at strong coupling, and we will observe that this precision is transferred to the defect case. Finally, if the bulk is fine-tuned to criticality, RCMPS no longer provide as good an approximation of long distance physics, and introduce an effective length-scale associated to the limited amount of entanglement they can represent. Precision is then reduced and extrapolations to the infinite entanglement limit \cite{pollmann2009_finiteentanglementscaling,vanhecke2022_finiteentanglementphi4} would be required. This is the first limitation of our method which makes it complementary to previous studies that considered scale invariant field theories in the bulk. 

 \begin{figure}[htp]
     \centering
    \begin{tikzpicture}

    \begin{scope}[xshift=-2.5cm]
        \node at (-1,2.0) {a)};
          
        \draw[->] (0,0) -- (0,2.3) node[left] {$\tau$};
        \draw[->] (0,0) -- (3,0) node[below] {$x$};  
        
        \draw[thick,blue] (-0.4,0.5) -- (0.8,0.5);
        \draw[thick,blue] (1.5,0.5) -- (2.2,0.5);
        \draw[thick,blue,dashed] (2.3,0.5) -- (3,0.5);
        
        \cross{-0.2}{0.5}
        \cross{0.5}{0.5}
        \cross{1.2}{0.5}
        \cross{2}{0.5}
    
    \end{scope}
    
    \begin{scope}[xshift=2.5cm]
        \node at (-1,2.0) {b)};
          
        \draw[->] (0,0) -- (0,2.3) node[left] {$\tau$};
        \draw[->] (0,0) -- (3,0) node[below] {$x$};  
        
        \draw[thick,blue] (-0.4,0.5) -- (3,0.5);
        \cross{1}{1.2}
        \cross{2.5}{1.5}
    
    \end{scope}

\end{tikzpicture}
    \caption{a) an allowed configuration of defect (continuous line) and local operators (cross) that we can compute because all operator insertions are aligned b) a configuration that a plain variational method cannot deal with, because the operator insertions are not aligned with the defect. The latter setup could be accessed, in principle, with an evolution method like TDVP.}
    \label{fig:defect_allowed}
\end{figure}
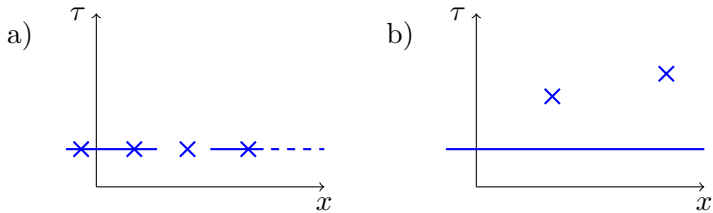

While the bulk model can be very general, we are limited in the type of defects we can study efficiently. First, the defects have to be linear in the field $\phi$, \ie correspond to operator insertions of the form $\exp(-\mu\int_0^L \phi) $. The geometry has flexibility: we allow combinations of defects of arbitrary length $L$ (even numerically semi-infinite or infinite) and strength $\mu$, along with arbitrary local operator insertions, \emph{as long as they are all aligned} (see Fig. \ref{fig:defect_allowed} for allowed configurations). This latter point is the second limitation of the method, and likely of any vanilla variational approach. This nonetheless includes impurity and complex local quench problems, as long as one is interested only in measurements on the impurity. In the local quench case, we remain limited to imaginary time, which is the third limitation of our approach at its present level of development.

\section{Setup}\label{sec:setup}
In this section, we explain how the $\phi^4$ model can be defined easily in the operator representation and then introduce a magnetic defect.

\subsection{A scalar field model without defects}

The $\phi^4$ model is (so far naively) defined by the Hamiltonian $H_{\text{bulk}}$
\begin{equation}\label{eq:Bulk_H}
    H_{\text{bulk}} := \int_\mathbb{R}  \frac{ \hpi^2}{2} + \frac{ (\nabla \hphi)^2}{2} +\frac{m^2}{2}\hphi^2 + g \hphi^4 = H_{m} + \int_\mathbb{R} g \hphi^4~\,,
\end{equation}
where $m$ is the bare mass and $g$ is the coupling constant. At this stage, this Hamiltonian is not yet well defined and one option to make sense of it is to first diagonalize the free part $H_m$ in terms normal modes $\hat{a}_k$. More explicitly, we expand the field $\hphi(x)$ and its conjugate momentum $\hpi(x)$
\begin{align}\label{eq:ModeExpansion}
    &\hphi(x) = \frac{1}{2\pi}\int \frac{\upd k}{\sqrt{2\omega_{k}}} (\hat{a}_{k}e^{ikx} + \hat{a}^{\dagger}_{k}e^{-ikx})\\
     &\hpi(x) = \frac{1}{2i\pi}\int \upd k \sqrt{\frac{\omega_k}{2}}(\hat{a}_{k}e^{ikx} - \hat{a}^{\dagger}_{k}e^{-ikx})~, \qquad \text{with} \quad [\hat{a}_{k},\hat{a}^{\dagger}_{k'}] = 2\pi \delta(k-k')~\,,
\end{align}
where $\hat{a}^{\dagger}_{k}$ and $\hat{a}_{k}$ are the creation and annihilation operators (respectively) of mode $k$, and $\omega_k = \sqrt{k^2+m^2}$. This yields: 
\begin{align}
    H_{m} &= \int_{\mathbb{R}}\frac{\upd k}{2\pi} \frac{\omega_{k}}{2}(\hat{a}_{k}\hat{a}^{\dagger}_{k} + \hat{a}^{\dagger}_{k}\hat{a}_{k})\,.
\end{align}
The Hamiltonian above describes infinitely-many decoupled harmonic oscillators, one for every value of $k$. The Hilbert space of this model is constructed by adding the excitations on top of the free vacuum $|0\rangle$ which is annihilated by $\hat{a}_{k}$'s. Normal ordering the free Hamiltonian $H_{m}$ w.r.t. mass $m$ sets the ground state energy density to zero. Now we have a legitimate Hamiltonian which is bounded from below, albeit without any interactions:
\begin{align}
    :H_{m}: = \int_{\mathbb{R}}\frac{\upd k}{2\pi} \omega_{k} \hat{a}^{\dagger}_{k}\hat{a}_{k}\,.
\end{align}
We can further apply this normal ordering to the interaction potential. In terms of Feynman diagrams, this normal ordering is equivalent to the suppression of tadpole graphs, which are the only divergences that appear in perturbation theory. In $d=1+1$, it has been shown in \cite{glimm1968_phi4bound,federbush1969_phi4bound} that normal ordering polynomial interactions is indeed sufficient beyond perturbation theory, and leads to a Hamiltonian \emph{density} that is bounded from below (at least as long as monomial of largest degree is even and with a positive coupling). Therefore, the $\phi^4$ Hamiltonian
\begin{align}\label{Hamiltonian_bulk}
    H_{\text{bulk}} &=~ \normord{H_{m}} +  \int_\mathbb{R}  g \normord{\hphi^4}~ = \int_\mathbb{R} h\,,
\end{align}
is amenable to the variational method: in the translation invariant case, we minimize a finite, lower-bounded scalar to get $\langle \Omega|h|\Omega\rangle$.

This model is arguably the simplest interacting QFT that is rigorously defined, while not being supersymmetric or integrable. As a result, it has been studied with various methods, each with its strengths and weaknesses: Hamiltonian truncation \cite{rychkov2015,Elias-Miro:2017xxf,elias-miro2017NLOrenormalization} (IR and energy cutoff), Monte-Carlo \cite{PhysRevD.79.056008,PhysRevD.99.034508,PhysRevD.92.034509} (both UV and IR cutoffs), MPS \cite{milsted2013_MPSforPhi4,PhysRevLett.123.250604} (UV cutoff), Tensor Network Renormalization \cite{Kadoh:2018tis,Delcamp:2020hzo} (UV cutoff), Borel Resummation \cite{Serone:2018gjo} (no cutoffs but perturbative). Aside from not being limited by any cutoff, RCMPS can straightforwardly handle \emph{any} interaction potential added to $H_{m}$ as long as the Hamiltonian density is bounded from below, with the same construction that is used for the $\hphi^{4}$ model. In this work, we restrict ourselves to the $\hphi^{4}$ model for simplicity and concreteness.

\subsection{Defects}
Next, we introduce a magnetic line defect of length $L$ in the model. In the Hamiltonian picture, this is a modification of the $\hphi^4$ Hamiltonian \eqref{Hamiltonian_bulk} where we turn on the additional interaction $-\mu \hphi(x_0)$ localized at $x_0=(x,0)$, for an interval of imaginary-time $x$ of length $L$. As explained in \ref{sec:Wick_Rotation}, upon exploiting Euclidean invariance of the $\phi^4$ model, we can view such modification as an extended operator
\begin{align}\label{defectop}
	\Dmu := \e^{-\mu \int_{-L}^0 \hphi}\,,
\end{align}
so that vacuum expectation values in this model can be computed as
\begin{equation}\label{correlators_gen}
   \langle \phi(x_1) \cdots \phi(x_n)\rangle_\text{defect} = \frac{\bra{0,g} \,\hphi(x_1) \cdots \hphi(x_n)\,  \Dmu \ket{0,g}}{\bra{0,g} \Dmu \ket{0,g}}\,,
\end{equation}
where $\ket{0,g}$ is the interacting vacuum of $H_\text{bulk}$ at coupling $g$. The defect expectation value is both UV and IR finite\footnote{By IR finite, we mean that the expectation value has a well defined limit when the defect size is fixed, but the thermodynamic limit is taken for the bulk, that is without global IR cutoff.} without normal ordering.

\section{Relativistic continuous matrix product states}\label{sec:RCMPSreview}

In this section we motivate the definition of RCMPS, explain how to compute the expectation value of local observables, quickly discuss how RCMPS are optimized to find the ground state of a quantum field theory, and finally show how extended operators (defects) can be evaluated. Apart from the last part about defects, this section is a summary of previous results \cite{Tilloy:2021hhb,Tilloy:2021yre,Tilloy:2022kcn}.

\subsection{A motivation} \label{sec:motivation}

Matrix product states are ansatz quantum states for spin chains (or chains of qudits).  They are parameterized non-linearly and in an extensive manner. Moreover, for the translation invariant case, the number of parameters does not grow with system size. They efficiently approximate states with a bounded amount of bipartite entanglement, such as the ground states of local gapped Hamiltonians. A natural strategy to solve a QFT like $\phi^4$ is thus to discretize it on a chain with lattice spacing $\varepsilon$, solve it with MPS, and finally numerically take the continuum limit $\varepsilon \rightarrow 0$ \cite{milsted2013_MPSforPhi4}. This works fairly well practically for fixed $\varepsilon$, but one cannot take $\varepsilon$ arbitrarily small. Indeed, a relativistic QFT, even massive, behaves like a critical theory at short distance. As a consequence, the amount of bipartite entanglement in the ground state of the lattice discretized model grows logarithmically with $\varepsilon^{-1}$. This makes the MPS approximation worse and worse as the continuum limit is approached (for a fixed number of variational parameters). Taking the continuum limit of MPS first analytically, with CMPS, does not bypass the divergent entanglement entropy problem. So while MPS do solve the IR problem admirably, they struggle with the short distance behavior that is typically found in relativistic QFT.

Hamiltonian truncation, on the other hand, has almost the opposite qualities. The idea is to first introduce a box of size $L$ (or equivalently an infrared cutoff), which has the effect of discretizing the momenta $k$ appearing in the normal modes $\hat{a}_k$. The resulting Hilbert space, spanned by all vectors of the form:
\begin{equation}\label{eq:HT_basis}
    \ket{k_1,m_1,k_2,m_2,\cdots k_n,m_n} := \frac{1}{\sqrt{m_1! \, m_2! \,\cdots m_n!}}a_{k_1}^{\dagger m_1} a_{k_2}^{\dagger m_2} \cdots a_{k_n}^{\dagger m_n} \ket{0}\,,
\end{equation}
where $\ket{0}$ is the free vacuum, is still an infinite dimensional Fock space, but with a discrete basis. One can further truncate it by considering only basis states below a certain cutoff energy $E_c$. Ultimately, one diagonalizes the Hamiltonian on the resulting finite dimensional Hilbert space to obtain the ground state. As far as short distance physics is concerned, this is an excellent scheme because the Hilbert space is constructed from excitations above $\ket{0}$. The latter, for super-renormalizable models, already contains an infinite amount of spatial entanglement, that diverges in exactly the same way as that of the true interacting ground state. However, just like exact diagonalization on the lattice, Hamiltonian truncation does not solve the many-body problem (or IR problem): for a fixed target error, the size of the truncated Hilbert space increases exponentially with system size. This is because, in contrast with matrix product state, the parameterization on the truncated Fock space is not extensive.

Ideally, one would like to benefit from the progress made both with HT and MPS and solve IR and UV difficulties together. This is doable: one should just construct a (continuous) matrix product state \emph{on top} of the free Fock vacuum. This combines the insight from HT that one should start from a spatially entangled state with the same (real space) divergence structure as the true ground state, with the intuition that MPS provide a good extensive parameterization adapted to the thermodynamic limit.

\subsection{The definition of RCMPS}
To construct a (continuous) MPS, one first needs to write the Hilbert space as a (continuous) tensor product of factors. Choosing factors associated to each normal mode would be natural, but not adapted because 1) physics is not translation invariant in momentum 2) the interacting ground state is not weakly entangled in momentum. Following \cite{Tilloy:2021yre,Tilloy:2021hhb}, we go back to real space and introduce $\hat{a}^{\dagger}(x)$ which is the Fourier transform of the canonical creation operators $\hat{a}^{\dagger}_k$:
\begin{equation}
    \hat{a}(x) = \frac{1}{2\pi } \int_\mathbb{R} \upd k \, \e^{ikx} \hat{a}_k \,.
\end{equation}
Note the absence of $\sqrt{\omega_k}$ factors in the denominator which are crucial to ensure that $a(x)$ has canonical commutation relation $[a(x),a^\dagger(y)] = \delta(x-y)$. This makes the Hilbert space a continuously infinite tensor product of factors associated with each oscillator at position $x$. This is all that is needed to define a CMPS.

Formally, a RCMPS on a line of size $L$ and with periodic boundary conditions is a quantum state parameterized by two $D\times D$ complex matrices $Q,R$ and defined as
\begin{equation}\label{eq:RCMPS}
	\ket{Q,R} = \tr \Big{\{}\mathcal{P} \exp \Big{[} \int_{0}^L \upd x \, Q\otimes \id + R\otimes \hat{a}^{\dagger}(x)\Big{]}\Big{\}}\ket{0}\,,
\end{equation}
where $\ket{0}$ is the free Fock vacuum verifying $\forall x $, $\hat{a}(x)\ket{0} = 0$, $\mathcal{P}\exp$ is the path-ordered exponential, and the trace is taken over the finite $D$-dimensional space on which $Q,R$ act. This representation \eqref{eq:RCMPS} makes the extensiveness of the state manifest, but it helps to expand\footnote{This expansion is done \eg in \cite{haegeman2013}. One option to obtain the result in a simple way is to introduce $R(x) := e^{Qx}\,R\, e^{-Qx}$, write $\ket{Q,R}$ as $\tr \Big{\{}\mathcal{P} \exp \Big{[} \int_{0}^L \upd x \, R(x)\, \hat{a}^{\dagger}(x)\Big{]}\Big{\}}\ket{0}$ and then Taylor expand the path-ordered exponential.} it explicitly in the Fock basis:
\begin{align}
    &\ket{Q,R} = \sum_{n=0}^{+\infty} \int_{0\leq x_1\leq ... \leq x_n \leq L} \upd x_1 \upd x_2 \dots \upd x_n \; \varphi_n(x_1,x_2,\dots,x_n) \;a^\dagger(x_1) \dots a^\dagger(x_n) \ket{0} \label{eq:RCMPS_decomposition}\\
    &\text{with} ~~~ \varphi_n(x_1,x_2,\dots,x_n) := \tr\left[\e^{x_1 Q} R\,  e^{(x_2-x_1) Q} R \dots e^{(x_n-x_{n-1})Q} R \, e^{(L-x_n) Q} \right]\,. \label{eq:RCMPS_wavefunction}
\end{align}
This shows that while the RCMPS has few parameters, it is \emph{not} sparse in the Fock space, and all the basis coefficients are generically non-zero. The reader unfamiliar with MPS techniques may very well forget the motivation \ref{sec:motivation} of RCMPS and take its expanded form \eqref{eq:RCMPS_decomposition} as starting point.

We now make a few remarks about the RCMPS we just defined. First, the parameterization with $Q$ and $R$ is redundant. The state $\ket{Q,R}$ is invariant under $Q \xrightarrow{} X^{-1}QX$ and $R \xrightarrow{} X^{-1}RX$. We can exploit this freedom to impose a gauge condition $Q + Q^{\dagger} + R^{\dagger}R\, = \,0$, the so-called \emph{left canonical gauge}. With this choice, which we make from now on, one can write $Q= -i K -R^\dagger R/2$ with $K$ self-adjoint. Second, and with this choice of gauge, one can take the thermodynamic limit seamlessly, because the state has constant norm in the infinite size limit. Thus, at least within correlation functions, one can replace the integration interval $[0,L]$ in \eqref{eq:RCMPS} by $\mathbb{R}$.

Finally, we note that the field operator $\hphi(x)$ is not a local function of $a(x)$ and $a^\dagger(x)$, but is rather obtained from a convolution (see \eqref{eq:PhiIna} below). This is where RCMPS differ from standard CMPS, where one would take oscillator modes $\hat{\psi}(x)$ local in the field, \ie with $\hat{\psi}^{\dagger}(x) \propto \sqrt{\Lambda}\hphi(x) - i\hpi(x)/{\sqrt{\Lambda}}$ with some arbitrary UV scale $\Lambda$, and replace $\ket{0}$ by the state $\ket{0}_\psi$ annihilated by all the $\psi(x)$. While the Hamiltonian is stricly local in this representation, its ground state is infinitely entangled. RCMPS use a tensor factorization that breaks the strict locality of $H$, but yields a finitely entangled ground state.

\subsection{Computing local observables}\label{sec:RCMPS_local}

In many-body systems, and even more so in field theories, knowing a state does not necessarily imply that one can compute observables efficiently. We now show that, fortunately, we do get local observables at a cost $\propto D^3$ for RCMPS.

To this end, first note that since $\hat{a}^{\dagger}(x)$ satisfies the same commutation relation as $\hat{\psi}^{\dagger}(x)$, the results for expectation values for strings of $\hat{a}(x)$ follow from ``standard'' CMPS computations. 

In particular, normal-ordered correlation functions of products of $a$ and $a^\dagger$ can be computed by taking derivatives of the generating functional
\begin{equation}\label{eq:GeneratingRCMPS}
	\mathcal{Z}_{j',j} := \frac{\bra{Q,R}\exp[\int j'(x)\hat{a}^{\dagger}(x)]\exp[\int j(y)\hat{a}(y)]\ket{Q,R}}{\braket{Q,R|Q,R}}\,,
\end{equation}
with respect to the sources $j(x)$ and $j'(y)$. The generating functional admits a compact form in terms of $\mathbb{T}, j(x), j'(y)$ \cite{haegeman2013}:
\begin{align}\label{RCMPS_Generator}
	\mathcal{Z}_{j',j} & = \frac{\tr\left[\mathcal{P}\exp\left(\int \mathbb{T}+j R \otimes \id+j' \id \otimes R^*\right)\right]}{\tr\left[\mathcal{P}\exp\left(\int \mathbb{T}\right)\right]}\,,
\end{align}
where from now on we may assume the integrals are over $\mathbb{R}$. This is the central formula for CMPS from which every correlation function can in principle be derived.

Of course, we are interested in correlation functions of local operators such as $\hat{\phi}(x)$, not $\hat{a}(x)$. However, the latter can be obtained from those of $\hat{a}(x)$ and $\hat{a}^\dagger(x)$ through the convolution
\begin{equation}\label{eq:PhiIna}
	\hat{\phi}(x) = \int \upd y \, J(x-y)[\hat{a}(y) + \hat{a}^{\dagger}(y)]\,, \quad J(x) = \frac{1}{2\pi} \int \frac{\upd k}{\sqrt{2\omega_k}}e^{-i k x}~\,.
\end{equation}
This convolution is not too non-local: $J(x)$ has an integrable singularity at $x=0$, but decays exponentially as $|x|\rightarrow \infty$. It only introduces a mild inconvenience compared to the CMPS scenario but is not inhibitory in practice due to the fast decay of $J(x)$. In fact, it is precisely because we already pay the price of some non-locality in the RCMPS case that computing extended observables (or equivalently operators in the complete defect model) comes with no extra cost as we discuss in \ref{sec:RCMPS_defects}.

In principle, one could compute expectation values for polynomials of $:\hat{\phi}(x):$, by first computing expectation values of $\hat{a}(x)$ and then numerically convolving them with $J$. This would become increasingly costly as the degree of the polynomial increases. Fortunately, there is an easier way, which is to first compute the expectation value of a vertex operator $V_{b}(x) = :e^{b \hat{\phi}(x)}:$\, . The latter is indeed very simple in terms of the RCMPS generating functional $\mathcal{Z}_{j',j}$,
\begin{align}\label{eq:vertexops}
	\braket{V_{b}(0)}= \mathcal{Z}_{b J,b J}\,.
\end{align}
Now, we may obtain expectation values of powers of the field by differentiating $\braket{V_{b}(x)}$ w.r.t  $b$ and taking $b=0$ at the end. To compute $\mathcal{Z}_{bJ,bJ}$, we simply interpret the path ordered exponential as the solution of an ordinary differential equation (ODE):
\begin{align}
   &\mathcal{O}(L) := \mathcal{P}\exp \left(\int_0^L \mathbb{T}+j R \otimes \id+j' \id \otimes R^*\right)  ~~~\text{is indeed implied by}\\
   &\frac{\upd}{\upd x} \mathcal{O}(x) = \mathcal{O}(x) \, \left(\mathbb{T}+j R \otimes \id+j' \id \otimes R^*\right) ~~\text{and} ~~ \mathcal{O}(0) = \mathds{1} \,.
\end{align}
At this stage, it is customary to move to a superoperator representation $\mathbb{T} \mapsto \mathcal{L}$ to make explicit the fact that applying the map $\mathbb{T}$ scales as $D^3$ and not $D^4$ as it would naively if $\mathbb{T}$ were a generic map. Concretely, the superoperator representation is defined via the isomorphism $\ket{w}\otimes \ket{v} \cong \ket{w}\bra{v}$ \ie $\mathcal{L}$ is the representation of $\mathbb{T}$ acting on $D\times D$ matrices instead of vectors of lengths $D^2$. Explicitly:
\begin{align} 
&\mathbb{T}(\ket{w}\otimes \ket{v}) = Q\ket{w} \otimes \id\ket{v} +\id \ket{w} \otimes Q^*\ket{v} + R\ket{w}\otimes R^*\ket{v} \\
\implies &  \mathcal{L}(\ket{w}\bra{v}) = Q(\ket{w}\bra{v}) + (\ket{w}\bra{v})Q^{\dagger} + R(\ket{w}\bra{v})R^{\dagger}\,.
\end{align}
Hence, for a generic matrix $\rho$, 
\begin{align}\label{eq:Superop}
	\mathcal{L}\cdot\rho =  Q \,\rho + \rho \, Q^{\dagger} + R \, \rho \,  R^{\dagger}\,.
\end{align}
As a result, applying $\mathcal{L}$ can be done with simple matrix multiplications at cost $D^3$. Note that with the left-canonical gauge choice we made before, $\mathcal{L}$ is of the Lindblad form (and is thus in particular trace preserving).

Now combining the ODE and superoperator representations, we have $\langle V_b \rangle = \tr (\rho_\infty)$ with $\rho_\infty = \lim_{x\rightarrow\infty}\rho(x)$,
\begin{align}\label{eq:VertexODE}
	\frac{\upd}{\upd x} \rho(x) = \mathcal{L}\cdot \rho(x) +b J(x)[R \rho(x)+\rho(x)R^\dagger]\,,
\end{align}
and initial condition $\rho(-\infty) = \rho_0$, with $\mathcal{L}\cdot\rho_{0} = 0$ and $\tr[\rho_0] = 1$. 

To compute expectation values of powers of the field, we may now \emph{forward differentiate} this ordinary differential equation \eqref{eq:VertexODE} and evaluate the result at $b=0$. The kinetic part of the Hamiltonian density \eqref{eq:Bulk_H} is slightly more complicated to evaluate, since it also involves derivatives of the field and the conjugate momentum, but it can be computed with similar techniques and the same asymptotic cost \cite{Tilloy:2021yre}. 

\subsection{Optimizing the state}
We may therefore obtain the expectation value of the Hamiltonian density, $\braket{h}$, as the trace of a solution of linear matrix ODEs involving $Q$ and $R$. These ODEs can be evaluated to essentially arbitrary numerical precision using high-order Runge-Kutta solvers, with a cost dominated by the application of the generator, hence $D^3$. Further, using backpropagation (or equivalently adjoint methods) we can compute the full gradient of $\braket{h}$ with respect to the relevant parameters $K,R$ at the same cost $D^3$. 

We have everything needed to minimize the energy density over the parameters, and thus obtain a variational approximation to the ground state. The minimization could \emph{a priori} be done using any numerical solver (plain gradient descent or LBFGS). However, as expected from conventional tensor network wisdom, it was observed \cite{Tilloy:2021yre,Tilloy:2021hhb,Tilloy:2022kcn} that such naive minimizations work poorly, and get stuck in plateaus. The solution is to incorporate the geometry of RCMPS, that is, consider the metric induced on the parameters by the scalar product on the field theory Hilbert space. Taking into account this metric, which is different from the naive Euclidean metric, one can then perform the optimization on RCMPS states viewed as a Riemannian manifold. Crucially, the induced metric is efficiently computable and efficiently invertible, and thus Riemannian techniques are not slower per iteration than their naive Euclidean counterparts. Ultimately, one can even use Riemannian quasi-Newton methods (implemented \eg in \texttt{OptimKit.jl}) to further speed up convergence. The details of the numerical procedure can be found in \cite{Tilloy:2022kcn}.

\subsection{Introducing defects}\label{sec:RCMPS_defects}

We now explain how to introduce extended operators (or defects) within the previous formalism. Let us first consider the expectation value of the extended operator $\Dmu=\e^{-\mu \int_{-L}^0 \hphi}$. We recall that while the defect has a finite size $L$, we took the thermodynamic limit for the bulk. Given the RCMPS ground state $\ket{Q,R}$ of the $\phi^4$ model, the expectation value of $\Dmu$ is given by:
\begin{align}
	\langle 0,g |\Dmu |0,g\rangle \simeq \, \bra{Q,R} e^{-\mu \int_{-L}^0  \hphi}\ket{Q,R}\,.
\end{align}
Here, $\ket{Q,R} \simeq \ket{0,g}$ is the RCMPS approximation of a given bond dimension $D$ to the true ground state of the $\phi^4$ model at coupling $g$.  Using the Baker-Campbell-Hausdorff formula to normal order $\mathcal{D}_L$  (see appendix \ref{DefectExp} ) we have
\begin{align}\label{eq:defRCMPS}
	\begin{split}
		\bra{Q,R}\Dmu\ket{Q,R} = \mathcal{Z}_{-\mu G, -\mu G}\,\exp\left[\frac{\mu^{2}}{2}\int_{\mathbb{R} }\upd x\; G(x)^2\right]\,,
	\end{split}
\end{align}
with the modified ``source''
\begin{align}
	G(x) \equiv \int_{-L}^{0} \upd y \, J(x-y)\,,
\end{align}
and $J(x)$ being the source defined in eq.~\eqref{eq:PhiIna}. As before, the path ordered exponential in $ \mathcal{Z}_{-\mu G, -\mu G}$ is the solution to an ODE \eqref{eq:VertexODE} now involving the modified source $G(x)$. The additional exponential factor in $\braket{\mathcal{D}_{L}}_{Q,R}$ can be incorporated into the ODE as a term proportional to the identity to get (in superoperator form)
\begin{equation}\label{eq:denODE}
	\frac{\upd}{\upd x} \rho(x) = \mathcal{L}\cdot \rho(x) - \mu G(x) \left[ R\rho(x) + \rho(x) R^\dagger \right] + \frac{\mu^{2}}{2} G^{2}(x) \rho(x)\,,
\end{equation}
which gives
\begin{align}
	\bra{Q,R}\Dmu\ket{Q,R} = \lim_{x \to \infty} \, \mathrm{tr}\,[\rho(x)]~, ~~ \text{for the initial condition} \qquad \lim_{x\to -\infty} \rho(x)= \rho_{0}\,.
\end{align}
Hence, computing the expectation value of the defect operator, which is extended, is not more difficult (asymptotically in $D$) than computing expectation values of local operators, as we advertised.

Correlation functions of local operators in the defect theory can be computed in a similar manner. As in section \ref{sec:RCMPS_local}, we define the expectation value of a vertex operator in the full defect theory as
\begin{align}\label{eq:Defect_vertex}
    \langle V_b (x) \rangle_{\text{defect}} = \frac{\bra{0,g} :e^{b\hat{\phi(x)}}: \text{e}^{-\mu \int_{-L}^{0} \hat{\phi}}\ket{0,g}}{\bra{0,g}\text{e}^{-\mu \int_{-L}^{0} \hat{\phi}}\ket{0,g}}~\,,
\end{align}
where $\ket{0,g}$ is the exact ground state of the Hamiltonian at coupling $g$.
This expression, upon iterated differentiation with respect to $b$, allows one to compute expectation value of $:\phi^k(x):$ in the defect theory. We just computed the denominator, and the numerator can be evaluated in a similar way (see \ref{sec:DefectVertex} for more details). Ultimately, we obtain 
\begin{align}
	\begin{split}
		\frac{\bra{Q,R} V_{b}(x) \Dmu \ket{Q,R}}{\bra{Q,R} \Dmu \ket{Q,R}}
		& = \frac{\mathcal{Z}_{s_x,s_x}}{\mathcal{Z}_{-\mu G, -\mu G}} \; \times\exp\Big{(}-b\,\mu\int_{\mathbb{R}} \upd y \int_{-L}^{0} \upd z \:J(x-y)J(y-z)\Big{)}\,,
	\end{split} \label{eq:vertex_defect_solution}
\end{align} where
\begin{equation}
	s_x(y) = b\,J(x-y) - \mu\int_{-L}^{0} \upd z \: J(z-y)\,.
\end{equation}
Again, this means that the expectation value of the vertex operator in the defect theory can be computed efficiently as the ratio of the trace of the solution of matrix ODEs. Numerically, the only new difficulty is that r.h.s. of \eqref{eq:vertex_defect_solution} is the ratio of two terms that grow exponentially in $L$, the defect size. This can however be alleviated with simple normalization strategies outlined in appendix \ref{app:normalization_tricks}, which allow one to take the large defect limit $L\rightarrow \infty$ numerically.

\section{Results}\label{sec:results}

We can now numerically compute the expectation values of the defect operator $\Dmu$ and the field expectation value $\phi(x)$ with RCMPS for various values of the couplings $g$ and $\mu$. Since the physics only depends on $g/m^2$ and $\mu/m^2$, we fix $m=1$ in all the RCMPS simulations, without loss of generality. First, we recall some perturbation theory basics that allow us to benchmark the RCMPS results in the weak coupling regime $g=0.1$. Then, we move to the strong coupling regime, $g=2$, where we can only compare RCMPS with itself. Subsequently, we increase the coupling to its critical value, $g_c \simeq 2.771525$ \cite{Delcamp:2020hzo,vanhecke2022_finiteentanglementphi4}, a regime that is particularly difficult with RCMPS at fixed bond dimension, but allows us to have some comparisons with conformal field theory results. Finally, we push the coupling to the symmetry-broken phase, $g=4$, a regime where RCMPS allow us to consider two fundamentally different types of symmetry breaking: a natural one, following the defect perturbation, and a ``frustrated'' one, where the bulk is symmetry broken in the opposite way as the defect.

\subsection{Perturbation theory calculations}
Perturbation theory for the plain $\phi^4$ model (without the defect) is standard and explained in most textbooks \cite{Bjorken:1965zz, Peskin:1995ev}. An advantage of the $1+1$ dimensional case is that only the tadpole graphs are divergent, and they are removed by normal-ordering. Hence, all the Feynman diagrams appearing in the expansion are finite, and can be computed numerically \eg with Monte Carlo \cite{Lepage:2020tgj}.

To understand what happens when we add a defect, we first consider the free bulk case, \ie $g=0$. Since the defect is linear in the field, the whole functional integral is still Gaussian and one can compute it exactly. But, anticipating the interacting case, it still helps to see the result diagrammatically

For the defect expectation value in the free model, \ie $g=0$ and any $\mu$, we have that:
\begin{align}
\bra{0,0} \Dmu \ket{0,0} =1+\sum_{k>0} \frac{ (-\mu)^k}{k!}\bra{0,0}  \left(\prod_{j=1}^k \int_{-L}^0 \upd x_j \hphi(x_j)\right) \ket{0,0}\,,
\end{align}
where $\ket{0,0}$ is the ground state at $g=0$, \ie the free Fock vacuum of $H_m$.
By systematically applying Wick's theorem, we can rewrite the above as a sum of products of integrated free massive propagators. Explicitly, to all orders in $\mu$ we have that:
\begin{align}\label{vevDeffree0}
\bra{0,0} \Dmu \ket{0,0} &=1+\frac{\mu^2}{2} \int_{-L}^0 \int_{-L}^0 \upd x_1  \upd x_2 \, G(x_1-x_2;m) +\dots\nonumber\\
	\;\; &=\exp \Big{[}\frac{\mu^2}{2}   \int_{-L}^0 \int_{-L}^0 \upd x_1  \upd x_2 \,G(x_1-x_2;m)\Big{]}\Big{\}}\,,
\end{align}
with the free massive propagator given by
\begin{align}\label{propagator}
	G(x_i-x_j;m):= \bra{0,0} \hphi(x_i) \hphi(x_j) \ket{0,0}&=\frac{1}{2 \pi}  K_0\left(m |x_{i}-x_j|\right)\,,
\end{align}
where $K_0$ is the modified Bessel function of the second kind. In terms of the following Feynman rules
\begin{align}
	\begin{array}{c}
		\begin{tikzpicture}[baseline,valign]
			\draw[thick] (0, 0) -- (1.4, 0);
			\node[below] at (0, 0) {$x_1$};
			\node[below] at (1.4, 0) {$x_2$};
		\end{tikzpicture}
	\end{array}
	\;\; &\equiv \;\;
	G(x_1,x_2; m)\,, \nn\\
	\begin{array}{c}
		\begin{tikzpicture}[baseline,valign]
			\draw[dashed, blue] (0, 0) -- (1.4, 0);
			\draw[thick] (0.7, 0) to (0.7, 0.7);
			\node at (0.7, 0.0) [dcirc] {};
		\end{tikzpicture}
	\end{array}
	\;\;~~&\equiv \;\;
	-\mu  \int_{-L}^0 \upd x\,,
\end{align}
we can rewrite \eqref{vevDeffree0} as
\begin{align}\label{vevDeffree}
\bra{0,0} \Dmu \ket{0,0} 	 & \;\;=
	\begin{tikzpicture}[baseline,valign]
		\draw[dashed, blue] (0, 0) -- (1.3, 0);
	\end{tikzpicture} 	\;\; + \;\;
	\tfrac{1}{2}\begin{tikzpicture}[baseline,valign]
		\draw[dashed, blue] (0, 0) -- (1.3, 0);
		\draw[thick] (0.2, 0) to[out=90,in=90] (1.1, 0);
		\node at (1.1, 0.0) [dcirc] {};
		\node at (0.2, 0.0) [dcirc] {};
	\end{tikzpicture}
	\;\; + \;\;
	\tfrac{1}{8}\begin{tikzpicture}[baseline,valign]
		\draw[dashed, blue] (0, 0) -- (1.3, 0);
		\draw[thick] (0.2, 0) to[out=90,in=90] (1.1, 0);
		\draw[thick] (0.1, 0) to[out=90,in=90] (1.2, 0);
		\node at (1.2, 0.0) [dcirc] {};
		\node at (1.1, 0.0) [dcirc] {};
		\node at (0.1, 0.0) [dcirc] {};
		\node at (0.2, 0.0) [dcirc] {};
	\end{tikzpicture}
	\;\; + \;\;
	\tfrac{1}{48}\begin{tikzpicture}[baseline,valign]
		\draw[dashed, blue] (0, 0) -- (1.3, 0);
		\draw[thick] (0.2, 0) to[out=90,in=90] (1.1, 0);
		\draw[thick] (0.1, 0) to[out=90,in=90] (1.2, 0);
		\draw[thick] (0.0, 0) to[out=90,in=90] (1.3, 0);
		\node at (1.2, 0.0) [dcirc] {};
		\node at (1.1, 0.0) [dcirc] {};
		\node at (0.1, 0.0) [dcirc] {};
		\node at (0.2, 0.0) [dcirc] {};
		\node at (0.0, 0.0) [dcirc] {};
		\node at (1.3, 0.0) [dcirc] {};
	\end{tikzpicture}
	\;\; +\dots  \;\;\nn\\
	\;\; & \;\;=\exp \Big{[} 	\tfrac{1}{2}\begin{tikzpicture}[baseline,valign]
		\draw[dashed, blue] (0, 0) -- (1.3, 0);
		\draw[thick] (0.2, 0) to[out=90,in=90] (1.1, 0);
		\node at (1.1, 0.0) [dcirc] {};
		\node at (0.2, 0.0) [dcirc] {};
	\end{tikzpicture} \Big{]}\,.
\end{align}

For the interacting model with $g\neq 0$ we proceed similarly, expanding both in $g$ and $\mu$.  This time, the Feynman rules include bulk vertices
\begin{align}
	\begin{tikzpicture}[baseline,valign]
		\draw[thick] (0.5, 0.5) -- (-0.5, -0.5);
		\draw[thick] (0.5, -0.5) -- (-0.5, 0.5);
		\node at (0, 0.0) [rcirc] {};
	\end{tikzpicture}
	\;\; &\equiv \;\;
	-4!g \int d^2 x
	\,,
\end{align}
as well as propagators connecting the bulk and points on the defect. These propagators are given in eq.~\eqref{propagator}, where now each point can be in the bulk or on the defect.

At any fixed order in $g$, there are infinitely-many ``disconnected'' diagrams (\ie where bulk and defect vertices are not connected to each other by any propagator), that ``dress'' the defect identity by defect interactions, just as in \eqref{vevDeffree}. At any fixed order in $g$, each connected diagram is ``dressed'' by the same factor, and there are \emph{finitely-many} connected diagrams. In other words, and as far as connected correlators are considered, at any fixed order in $g$ the $\mu$-expansion truncates. Crucially, as we will later observe, this does not imply that the expansion is accurate at a fixed small $g$ and arbitrarily large $\mu$!

For the vacuum expectation value of $\Dmu$ we find (neglecting $\order(g^5)$ corrections):
\begin{align}\label{defect_vev_pert}
\frac{	\bra{0,g} \Dmu \ket{0,g}}{\bra{0,0} \Dmu \ket{0,0}} &= 1+
	\tfrac{1}{24}\begin{tikzpicture}[baseline,valign]
		\draw[dashed, blue] (-0.4, 0) -- (1.4, 0);
		\draw[thick] (-0.2, 0) -- (0.5, 0.7) -- (1.2, 0);
		\draw[thick] ( 0.3, 0) -- (0.5, 0.7) -- (0.7, 0);
		\node at (0.5, 0.7) [rcirc] {};
		\node at (0.3, 0.) [dcirc] {};
		\node at (-0.2, 0.) [dcirc] {};
		\node at (1.2, 0.) [dcirc] {};
		\node at (0.7, 0.) [dcirc] {};
	\end{tikzpicture}
	+\tfrac{1}{12}\begin{tikzpicture}[baseline,valign]
		\draw[dashed, blue] (-0.4, 0) -- (1.4, 0);
		\draw[thick] (0.2, 0.6) -- (0.8, 0.6);
		\draw[thick] (0.2, 0.6) to[out=90,in=90] (0.8, 0.6);
		\draw[thick] (0.2, 0.6) to[out=-90,in=-90] (0.8, 0.6);
		\draw[thick] (0.8, 0.6) -- (0.8, 0);
		\draw[thick] (0.2, 0.6) -- (0.2, 0);
		\node at (0.2, 0.) [dcirc] {};
		\node at (0.8, 0.) [dcirc] {};
		\node at (0.2, 0.6) [rcirc] {};
		\node at (0.8, 0.6) [rcirc] {};
	\end{tikzpicture}
	+\tfrac{1}{16}\begin{tikzpicture}[baseline,valign]
		\draw[dashed, blue] (-0.4, 0) -- (1.4, 0);
		\draw[thick] (0.2, 0.6) to[out=90,in=90] (0.8, 0.6);
		\draw[thick] (0.2, 0.6) to[out=-90,in=-90] (0.8, 0.6);
		\draw[thick] (0.8, 0.6) -- (0.8, 0);
		\draw[thick] (0.8, 0.6) -- (1.2, 0);
		\draw[thick] (0.2, 0.6) -- (0.2, 0.0);
		\draw[thick] (0.2, 0.6) -- (-0.2, 0.0);
		\node at (0.2, 0.) [dcirc] {};
		\node at (1.2, 0.) [dcirc] {};
		\node at (-0.2, 0.) [dcirc] {};
		\node at (0.8, 0.) [dcirc] {};
		\node at (0.2, 0.6) [rcirc] {};
		\node at (0.8, 0.6) [rcirc] {};
	\end{tikzpicture}
	+\tfrac{1}{72}\begin{tikzpicture}[baseline,valign]
		\draw[dashed, blue] (-0.4, 0) -- (1.4, 0);
		\draw[thick] (0.2, 0.6) -- (0.8, 0.6);
		\draw[thick] (0.8, 0.6) -- (0.8, 0);
		\draw[thick] (0.8, 0.6) -- (1.2, 0);
		\draw[thick] (0.8, 0.6) -- (1.0, 0);
		\draw[thick] (0.2, 0.6) -- (0, 0);
		\draw[thick] (0.2, 0.6) -- (0.2, 0.0);
		\draw[thick] (0.2, 0.6) -- (-0.2, 0.0);
		\node at (0.2, 0.) [dcirc] {};
		\node at (0., 0.) [dcirc] {};
		\node at (1.2, 0.) [dcirc] {};
		\node at (-0.2, 0.) [dcirc] {};
		\node at (0.8, 0.) [dcirc] {};
		\node at (1, 0.) [dcirc] {};
		\node at (0.2, 0.6) [rcirc] {};
		\node at (0.8, 0.6) [rcirc] {};
	\end{tikzpicture}\nn\\
	&+
	\tfrac{1}{24}\begin{tikzpicture}[baseline,valign]
		\draw[dashed, blue] (-0.4, 0) -- (1.4, 0);
		\draw[thick] (0.2, 0.6) to[out=-30,in=-150] (0.8, 0.6);
		\draw[thick] (0.2, 0.6) to[out=30,in=150] (0.8, 0.6);
		\draw[thick] (0.2, 0.6) to[out=90,in=90] (0.8, 0.6);
		\draw[thick] (0.2, 0.6) to[out=-90,in=-90] (0.5, 0.4);
		\draw[thick] (0.5, 0.4) to[out=-90,in=-90] (0.8, 0.6);
		\draw[thick] (0.5, 0.35) -- (0.8, 0);
		\draw[thick] (0.5, 0.35) -- (0.2, 0);
		\node at (0.2, 0.) [dcirc] {};
		\node at (0.8, 0.) [dcirc] {};
		\node at (0.2, 0.6) [rcirc] {};
		\node at (0.8, 0.6) [rcirc] {};
		\node at (0.5, 0.35) [rcirc] {};
	\end{tikzpicture}
	+\tfrac{1}{8}\begin{tikzpicture}[baseline,valign]
		\draw[dashed, blue] (-0.4, 0) -- (1.4, 0);
		\draw[thick] (0.2, 0.6) to[out=90,in=90] (0.8, 0.6);
		\draw[thick] (0.2, 0.6) to[out=-90,in=-90] (0.5, 0.4);
		\draw[thick] (0.5, 0.4) to[out=-90,in=-90] (0.8, 0.6);
		\draw[thick] (0.2, 0.6) to[out=90,in=90] (0.5, 0.4);
		\draw[thick] (0.5, 0.4) to[out=90,in=90] (0.8, 0.6);
		\draw[thick] (0.8, 0.6) -- (0.8, 0);
		\draw[thick] (0.2, 0.6) -- (0.2, 0);
		\node at (0.2, 0.) [dcirc] {};
		\node at (0.8, 0.) [dcirc] {};
		\node at (0.2, 0.6) [rcirc] {};
		\node at (0.8, 0.6) [rcirc] {};
		\node at (0.5, 0.35) [rcirc] {};
	\end{tikzpicture}
	+\tfrac{1}{32}\begin{tikzpicture}[baseline,valign]
		\draw[dashed, blue] (-0.4, 0) -- (1.4, 0);
		\draw[thick] (0.2, 0.6) to[out=-90,in=-90] (0.5, 0.4);
		\draw[thick] (0.5, 0.4) to[out=-90,in=-90] (0.8, 0.6);
		\draw[thick] (0.2, 0.6) to[out=90,in=90] (0.5, 0.4);
		\draw[thick] (0.5, 0.4) to[out=90,in=90] (0.8, 0.6);
		\draw[thick] (0.8, 0.6) -- (0.8, 0);
		\draw[thick] (0.8, 0.6) -- (1.2, 0);
		\draw[thick] (0.2, 0.6) -- (0.2, 0.0);
		\draw[thick] (0.2, 0.6) -- (-0.2, 0.0);
		\node at (0.2, 0.) [dcirc] {};
		\node at (1.2, 0.) [dcirc] {};
		\node at (-0.2, 0.) [dcirc] {};
		\node at (0.8, 0.) [dcirc] {};
		\node at (0.2, 0.6) [rcirc] {};
		\node at (0.8, 0.6) [rcirc] {};
		\node at (0.5, 0.35) [rcirc] {};
	\end{tikzpicture}
	+\tfrac{1}{8}\begin{tikzpicture}[baseline,valign]
		\draw[dashed, blue] (-0.4, 0) -- (1.4, 0);
		\draw[thick] (0.2, 0.6) -- (0.8, 0.6);
		\draw[thick] (0.2, 0.6) to[out=90,in=90] (0.8, 0.6);
		\draw[thick] (0.2, 0.6) to[out=-90,in=-90] (0.5, 0.4);
		\draw[thick] (0.5, 0.4) to[out=-90,in=-90] (0.8, 0.6);
		\draw[thick] (0.5, 0.35) -- (0.8, 0);
		\draw[thick] (0.5, 0.35) -- (0.2, 0);
		\draw[thick] (-0.2, 0.0) -- (0.2, 0.6);
		\draw[thick] (1.2, 0.0) -- (0.8, 0.6);
		\draw[thick] (0.5, 0.35) -- (0.8, 0);
		\node at (0.2, 0.) [dcirc] {};
		\node at (1.2, 0.) [dcirc] {};
		\node at (-0.2, 0.) [dcirc] {};
		\node at (0.8, 0.) [dcirc] {};
		\node at (0.2, 0.6) [rcirc] {};
		\node at (0.8, 0.6) [rcirc] {};
		\node at (0.5, 0.35) [rcirc] {};
	\end{tikzpicture} \nn\\
	&+\tfrac{1}{36}\begin{tikzpicture}[baseline,valign]
		\draw[dashed, blue] (-0.4, 0) -- (1.4, 0);
		\draw[thick] (0, 0.6) -- (0.8, 0.6);
		\draw[thick] (0.2, 0.6) to[out=90,in=90] (0.8, 0.6);
		\draw[thick] (0.2, 0.6) to[out=-90,in=-90] (0.8, 0.6);
		\draw[thick] (0.8, 0.6) -- (0.8, 0);
		\draw[thick] (0, 0.6) -- (0.2, 0);
		\draw[thick] (0, 0.6) -- (-0.2, 0);
		\draw[thick] (0, 0.6) -- (0, 0);
		\node at (0.2, 0.) [dcirc] {};
		\node at (0, 0.) [dcirc] {};
		\node at (-0.2, 0.) [dcirc] {};
		\node at (0.8, 0.) [dcirc] {};
		\node at (0.2, 0.6) [rcirc] {};
		\node at (0.8, 0.6) [rcirc] {};
		\node at (0, 0.6) [rcirc] {};
	\end{tikzpicture}
	+\tfrac{1}{24}\begin{tikzpicture}[baseline,valign]
		\draw[dashed, blue] (-0.4, 0) -- (1.4, 0);
		\draw[thick] (0, 0.6) -- (0.4, 0.6);
		\draw[thick] (0.4, 0.6) to[out=90,in=90] (1, 0.6);
		\draw[thick] (0.4, 0.6) to[out=-90,in=-90] (1, 0.6);
		\draw[thick] (1, 0.6) -- (1, 0);
		\draw[thick] (0, 0.6) -- (0.2, 0);
		\draw[thick] (0, 0.6) -- (-0.2, 0);
		\draw[thick] (0, 0.6) -- (0, 0);
		\draw[thick] (0.4, 0.6) -- (0.4, 0);
		\draw[thick] (1, 0.6) -- (1.2, 0);
		\node at (1.2, 0.) [dcirc] {};
		\node at (0.2, 0.) [dcirc] {};
		\node at (0.4, 0.) [dcirc] {};
		\node at (0, 0.) [dcirc] {};
		\node at (-0.2, 0.) [dcirc] {};
		\node at (1, 0.) [dcirc] {};
		\node at (0.4, 0.6) [rcirc] {};
		\node at (1, 0.6) [rcirc] {};
		\node at (0, 0.6) [rcirc] {};
	\end{tikzpicture}
	+\tfrac{1}{48}\begin{tikzpicture}[baseline,valign]
		\draw[thick] (0.4, 0.6) to[out=90,in=90] (-0.5, 0);
		\draw[thick] (0.4, 0.6) to[out=90,in=90] (1.3, 0);
		\draw[dashed, blue] (-0.5, 0) -- (1.5, 0);
		\draw[thick] (0.0, 0.25) -- (0.8, 0.25);
		\draw[thick] (0.0, 0.25) -- (0.4, 0.6)  -- (0.8, 0.25);
		\draw[thick] (0, 0.25) -- (-0.25, 0);
		\draw[thick] (0, 0.25) -- (0.25, 0);
		\draw[thick] (0.8, 0.25) -- (0.6, 0);
		\draw[thick] (0.8, 0.25) -- (1, 0);
		\node at (-0.25, 0.) [dcirc] {};
		\node at (0.25, 0.) [dcirc] {};
		\node at (0.6, 0.) [dcirc] {};
		\node at (1, 0.) [dcirc] {};
		\node at (0.4, 0.6) [rcirc] {};
		\node at (0, 0.25) [rcirc] {};
		\node at (0.8, 0.25) [rcirc] {};
		\node at (-0.5, 0.) [dcirc] {};
		\node at (1.3, 0.) [dcirc] {};
	\end{tikzpicture}
	+\tfrac{1}{144}\begin{tikzpicture}[baseline,valign]
		\draw[dashed, blue] (-0.4, 0) -- (1.4, 0);
		\draw[thick] (0.2, 0.6) -- (0.8, 0.6);
		\draw[thick] (0.8, 0.6) -- (0.8, 0);
		\draw[thick] (0.8, 0.6) -- (1.2, 0);
		\draw[thick] (0.8, 0.6) -- (1.0, 0);
		\draw[thick] (0.2, 0.6) -- (0, 0);
		\draw[thick] (0.2, 0.6) -- (0.2, 0.0);
		\draw[thick] (0.2, 0.6) -- (-0.2, 0.0);
		\draw[thick] (0.5, 0.6) -- (0.4, 0.0);
		\draw[thick] (0.5, 0.6) -- (0.6, 0.0);
		\node at (0.2, 0.) [dcirc] {};
		\node at (0., 0.) [dcirc] {};
		\node at (0.6, 0.) [dcirc] {};
		\node at (0.4, 0.) [dcirc] {};
		\node at (1.2, 0.) [dcirc] {};
		\node at (-0.2, 0.) [dcirc] {};
		\node at (0.8, 0.) [dcirc] {};
		\node at (1, 0.) [dcirc] {};
		\node at (0.2, 0.6) [rcirc] {};
		\node at (0.5, 0.6) [rcirc] {};
		\node at (0.8, 0.6) [rcirc] {};
	\end{tikzpicture} \nn\\
	&+[34~\text{diagrams} ~\order{(g^4)}]\,.
\end{align}
For the one-point function we find:
\begin{align}\label{1pt_phi_pert}
\langle \phi(x) \rangle_\text{defect}&\equiv  
	\begin{tikzpicture}[baseline,valign]
		\draw[dashed, blue] (-.5, 0) -- (.5, 0);
		\draw[thick] (0, .8) -- (0, 0);
		\node at (0, 0) [dcirc] {};
	\end{tikzpicture}+\tfrac{1}{6}\begin{tikzpicture}[baseline,valign]
		\draw[dashed, blue] (-.5, 0) -- (.5, 0);
		\draw[thick] (0, .8) -- (0, 0);
		\draw[thick] (0, .4) -- (-.3, 0);
		\draw[thick] (0, .4) -- (.3, 0);
		\node at (0, 0) [dcirc] {};
		\node at (-.3, 0) [dcirc] {};
		\node at (.3, 0) [dcirc] {};
		\node at (0, .4) [rcirc] {};
	\end{tikzpicture}
	+\tfrac{1}{6}\begin{tikzpicture}[baseline,valign]
		\draw[dashed, blue] (-.5, 0) -- (.5, 0);
		\draw[thick] (0, .8) -- (0, 0);
		\draw[thick] (0, .2) to[out =0, in=0] (0, .6);
		\draw[thick] (0, .2) to[out=180,in=180] (0, .6);
		\node at (0, 0) [dcirc] {};
		\node at (0, .2) [rcirc] {};
		\node at (0, .6) [rcirc] {};
	\end{tikzpicture}
	+\tfrac{1}{4}\begin{tikzpicture}[baseline,valign]
		\draw[dashed, blue] (-.5, 0) -- (.5, 0);
		\draw[thick] (-.3, .5) -- (.2, .5);
		\draw[thick] (-.3, .5) -- (-.3, 0);
		\draw[thick] (.2, .5) -- (.3, 0);
		\draw[thick] (.2, .5) -- (.1, 0);
		\draw[thick] (-.3, .5) to[out =40, in=180-40] (.2, .5);
		\node at (-.3, .5) [rcirc] {};
		\node at (.2, .5) [rcirc] {};
		\node at (.1, .0) [dcirc] {};
		\node at (.3, .0) [dcirc] {};
		\node at (-.3, .0) [dcirc] {};
	\end{tikzpicture}
	+\tfrac{1}{12}\begin{tikzpicture}[baseline,valign]
		\draw[dashed, blue] (-.5, 0) -- (.5, 0);
		\draw[thick] (0, .8) -- (0, 0);
		\draw[thick] (0, .7) -- (-.4, 0);
		\draw[thick] (0, .4) -- (-.2, 0);
		\draw[thick] (0, .4) -- (.2, 0);
		\draw[thick] (0, .7) -- (.4, 0);
		\node at (0, 0) [dcirc] {};
		\node at (-.4, 0) [dcirc] {};
		\node at (-.2, 0) [dcirc] {};
		\node at (.2, 0) [dcirc] {};
		\node at (.4, 0) [dcirc] {};
		\node at (0, .4) [rcirc] {};
		\node at (0, .7) [rcirc] {};
	\end{tikzpicture}\nn\\
	&+[13~\text{diagrams} ~\order{(g^3)}]+[65~\text{diagrams} ~\order{(g^4)}]\,.
\end{align}
We compute the diagrams in the expansions \eqref{defect_vev_pert} and \eqref{1pt_phi_pert} up to order $g^4$ with an adaptive Monte Carlo method \cite{Lepage:2020tgj}, at an accuracy sufficient to make the integration error (presumably) far lower than the neglected order $g^5$ terms.

\subsection{\texorpdfstring{Weak coupling $g=0.1$, comparison with perturbation theory}{Weak coupling g=0.1, comparison with perturbation theory}}

\begin{figure}[htp]
	\centering
	\subfloat[]{
		\includegraphics[width=0.48\textwidth]{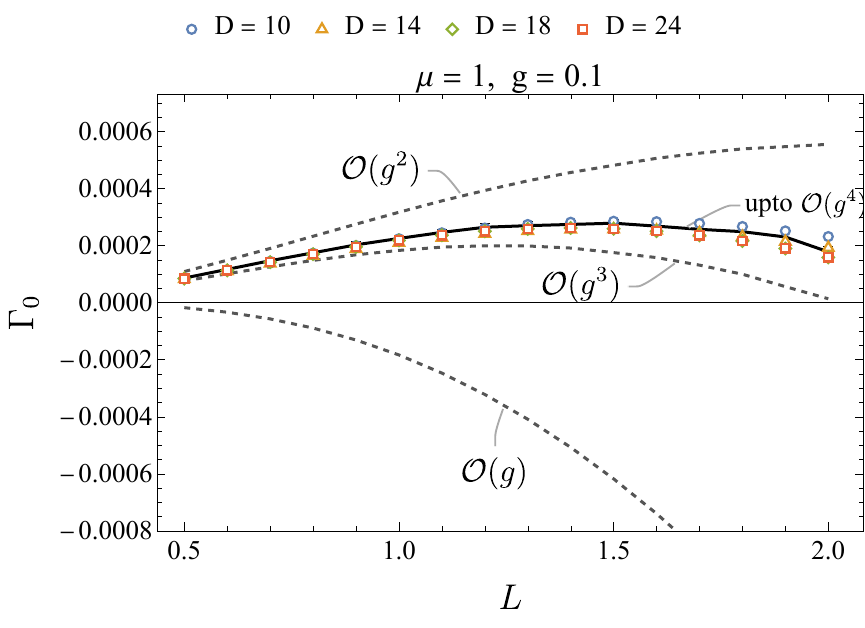}}
	\subfloat[]{
		\includegraphics[width=0.48\textwidth]{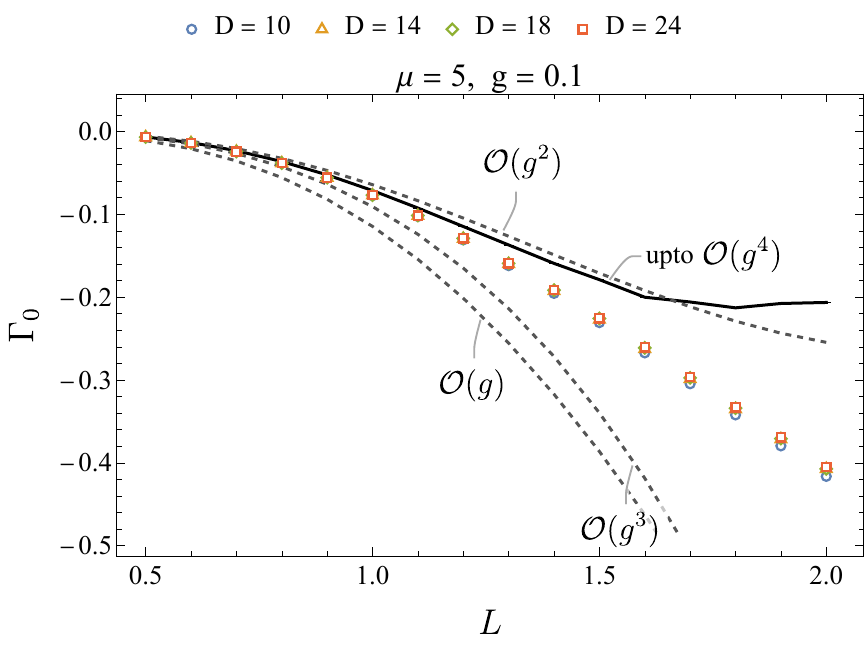}}
	\caption{Perturbative prediction for $\Gamma_0 \equiv  \frac{\bra{0,g} \Dmu \ket{0,g}}{\bra{0,0} \Dmu \ket{0,0}} -1$ up to $\mathcal{O}(g^4)$ vs. RCMPS (colored markers). Monte Carlo error bars are smaller than the size of the data points.}
	\label{fig:0pt_vs_L_pert}
\end{figure}

We first compare the perturbative calculation for the defect expectation value \eqref{defect_vev_pert} and RCMPS in Fig.~\ref{fig:0pt_vs_L_pert}, where we have chosen $g=0.1$ and considered two values $\mu=1$ and $\mu=5$ for the defect coupling. For $\mu=1$, we observe that a modest bond dimension is sufficient to get well-converged RCMPS results, whereas we have to go to a fairly high order, $\order{(g^4)}$, in the perturbation theory to get similar values. For larger $\mu$, the defect is no longer a small perturbation as $L$ grows, and the perturbation theory results are already fairly inaccurate at $\order{(g^4)}$ (at least without Borel-Padé resummation) while RCMPS results remain just as good.

\begin{figure}[htp]
	\centering
\subfloat[]{
	\includegraphics[width=0.48\textwidth]{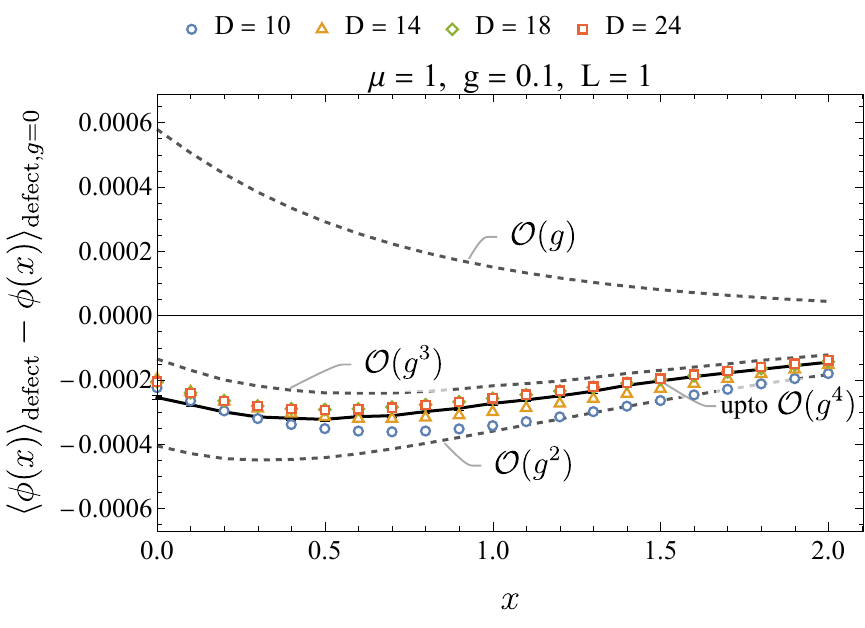}
}
\subfloat[]{
	\includegraphics[width=0.48\textwidth]{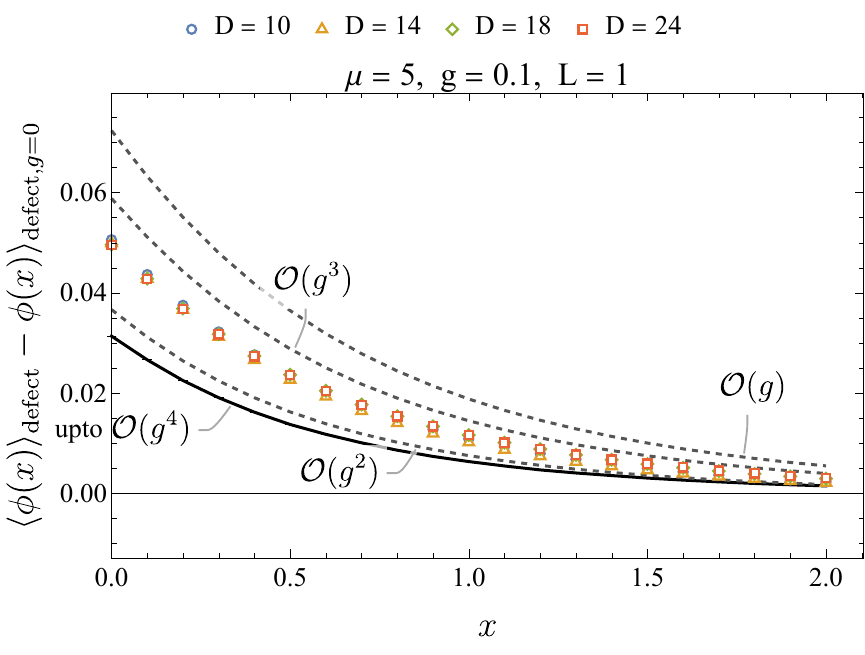}
}	
\caption{Up to $\order(g^4)$ perturbative prediction for $\equiv \langle \phi(x)\rangle _\text{defect} -\langle \phi(x)\rangle _{\text{defect}, g=0}$ vs. RCMPS (colored markers). Monte Carlo error bars are smaller than the thickness of the lines.}
	\label{fig:1pt_vs_tau_pert}
\end{figure}

Next we consider the one-point function $\langle\phi(x)\rangle_\text{defect}$, from which we subtract the free theory contribution, in Fig.~\ref{fig:1pt_vs_tau_pert}, for varying $x$, and fixed moderate defect size $L=1$. We make similar observations. First, at small $\mu$, RCMPS quickly converges, while only the highest order in perturbation theory quantitatively agrees. Second, at large $\mu$, perturbation theory does not seem to converge, but RCMPS results converge even faster.

From the fact that the $g$-expansion is \emph{exact} in $\mu$, which is a consequence of working with a defect that is linear in $\phi$, one might have expected perturbation theory to be accurate for fixed $g=0.1$ and large $\mu$. However, as is clear from these numerical results, this is not the case. In fact, from the Feynman diagram expansion, it is clear that for very small $g$, the expansion is inaccurate when $\mu$ is large. For example, for the case of defect expectation values in \eqref{defect_vev_pert}, we see that diagrams of order $n$ scale like $\order(g^n \mu^{2n+2})$, and thus become large when $\mu$ increases as $g$ is kept small but fixed. 

\subsection{\texorpdfstring{Strong coupling $g=2.0$ self-comparison}{Strong coupling g=2.0, self-comparison}}

\begin{figure}[htp]
    \centering
\subfloat[]{
	\includegraphics[width=0.48\textwidth]{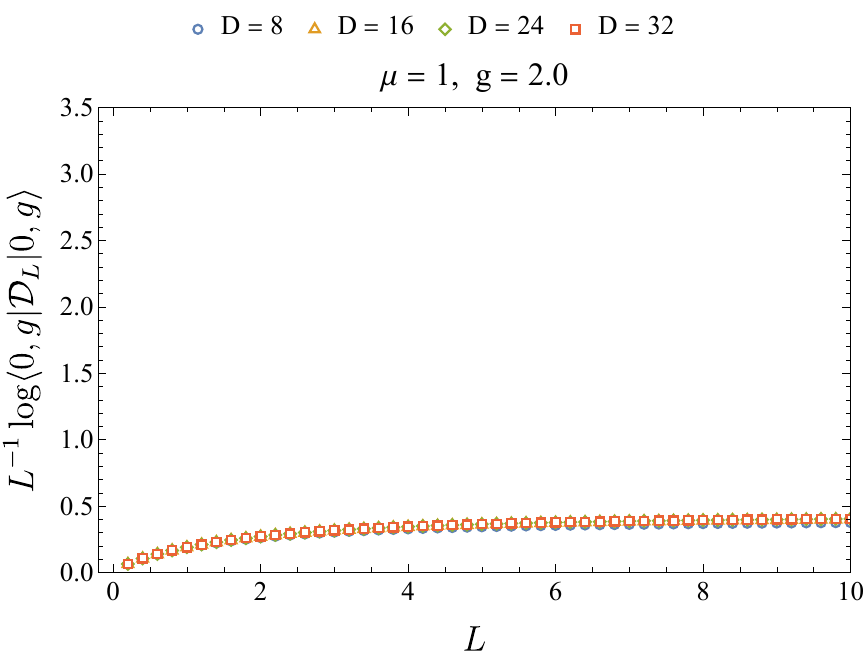}}
\subfloat[]{
	\includegraphics[width=0.48\textwidth]{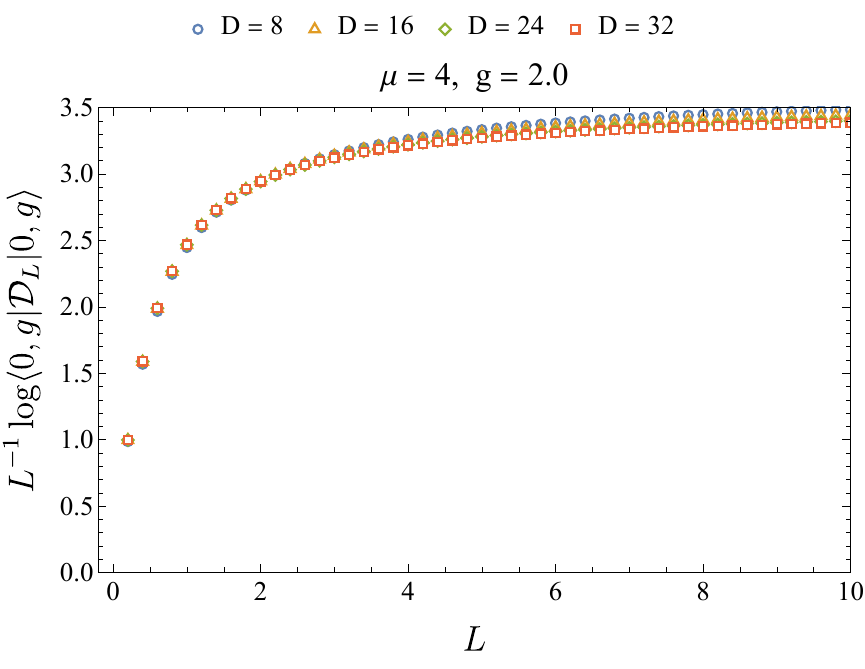}}
\caption{(Rescaled logarithm of the) Defect vacuum expectation value from RCMPS computation as a function of the defect length $L$ in the strong-coupling, symmetric regime for $\mu = 1$ (left) and $\mu=4$ (right).}
\label{fig:0pt_vs_L_strong}
\end{figure}

We now explore the strong-coupling (but still symmetric) regime $g=2$. 
Using RCMPS, we can compute the vacuum expectation value of $\Dmu$ for finite (and not necessary small) values of $g$ and $\mu$, and as a function of $L$. 
Since the defect expectation value grows exponentially as a function of $L$, we plot $L^{-1} \log\bra{0,g} \Dmu \ket{0,g}$ in Fig.~\ref{fig:0pt_vs_L_strong} for $\mu=1$ and $\mu=4$, which should go to a constant for large $L$.

As expected, the results converge quickly again, even for large $L$ and large $\mu$, in a regime where plain perturbation theory would be completely hopeless.

\begin{figure}[htp]
	\centering
	\subfloat[]{
		\includegraphics[width=0.48\textwidth]{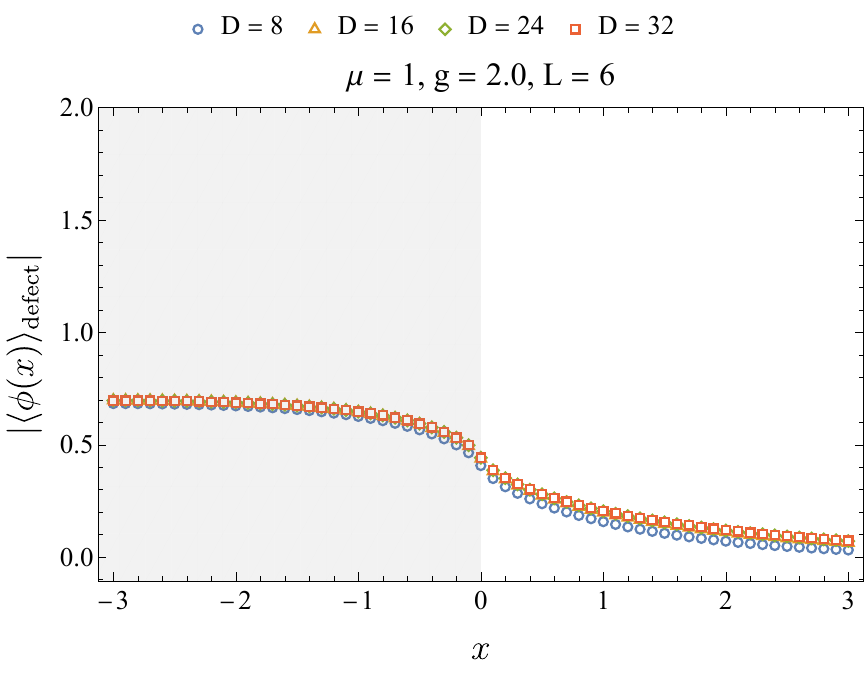}
	}
	\subfloat[]{
		\includegraphics[width=0.48\textwidth]{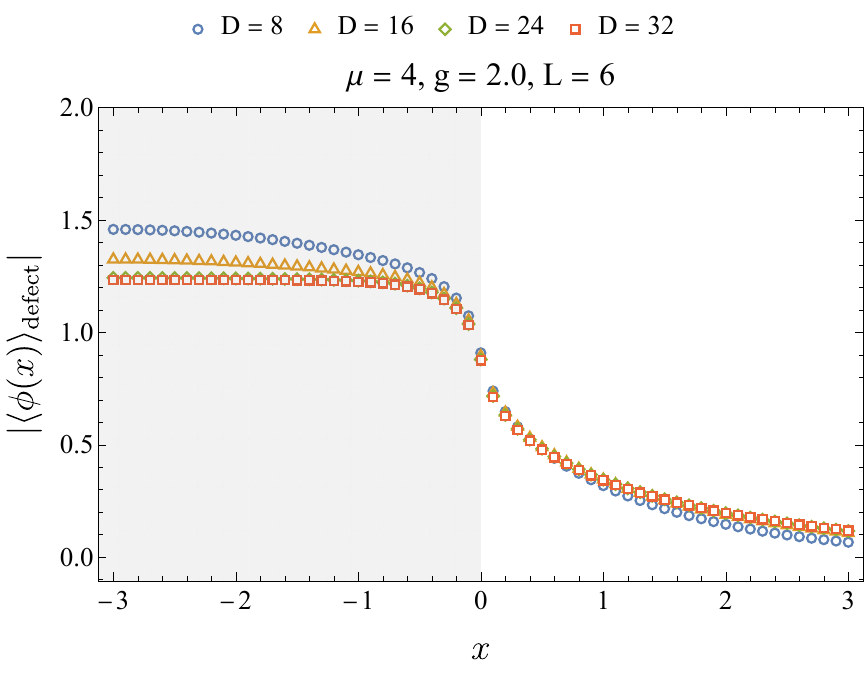}
	}	
	\caption{Bulk one-point functions from RCMPS computation as a function of the distance from the defect (shaded region) in the strong-coupling, symmetric regime, for $\mu=1$ (left) and $\mu=4$ (right).}
	\label{fig:1pt_vs_tau_strong}
\end{figure}

Next, we consider $|\langle \phi(x)\rangle_{\text{defect}}|$ for finite values of $g$, $\mu$ and $L$, as a function of $x$. For definiteness, in figure \ref{fig:1pt_vs_tau_strong} we show the result corresponding to the choice $g=2$, $\mu =1, 4$, and $L=6$.  The results agree with our physical intuition. First, Fig.~\ref{fig:1pt_vs_tau_strong} shows that $|\langle \phi(x)\rangle_{\text{defect}}|$ decays exponentially as we move away from the center of the defect  $x=-L/2$ into the bulk, as expected. Second, increasing the defect coupling $\mu$ (keeping all other parameters fixed), $|\langle \phi(x)\rangle_{\text{defect}}|$ increases, as it should since the bulk field tends to align with the defect. Third, the field expectation value remains continuous at $x=0$, \ie as we cross the defect. For $x<0$ it measures a \emph{defect magnetization} and stabilizes to a constant value in the middle of the defect.

At large $\mu$, the RCMPS results, although accurate at $D\geq 24$, are arguably slower to converge, especially inside the defect. We believe this is fairly intuitive, and a property of all variational methods. Our RCMPS states are approximate ground states of the bulk model, obtained by minimizing an energy density. From the ground state, one can define a probability distribution for the spatial average of the field in any finite size interval. The RCMPS approximates this probability very well overall, but is inevitably less accurate in the tails, for very small probabilities that almost do not contribute to the energy density. As we increase $\mu$ we shift the center of the distribution, and thus get more sensitive to its tail. As a result, we need a larger bond dimension to preserve the accuracy. This shift of the distribution is stronger closer to the defect, and even more so inside it, and thus we need larger bond dimensions inside the defect. Note however that even deep in the defect and at large $\mu$ we can easily reach bond dimensions such that the results are visually converged in Fig.~\ref{fig:1pt_vs_tau_strong}. We do not know of any other method that could reach a similar accuracy for such large perturbations both in $g$ and $\mu$ in the continuum.

\subsection{\texorpdfstring{Critical coupling $g\simeq 2.771525$, comparison with scaling expectations}{Critical coupling g = 2.771525, comparison with scaling expectations}}
We now tune the bulk model to its critical coupling, $g_c\simeq 2.771525$ \cite{Delcamp:2020hzo,vanhecke2022_finiteentanglementphi4}. In this strong coupling regime, one might be able to test RCMPS not just against itself, but also against universal scaling predictions from Conformal Field Theory. Indeed, at criticality, the $\phi^4$ model is in the Ising universality class, which is well understood. 

However, this universal regime is also where the accuracy of RCMPS degrades because of their finite entanglement entropy. In principle, one could improve the accuracy  of the RCMPS results by doing a finite entanglement scaling of the results \cite{pollmann2009_finiteentanglementscaling}, something we keep for future work. Here, we stick with a plain variational study, which also helps to demonstrate the limitations of RCMPS.

\begin{figure}[htp]
    \centering
\subfloat[]{
	\includegraphics[width=0.48\textwidth]{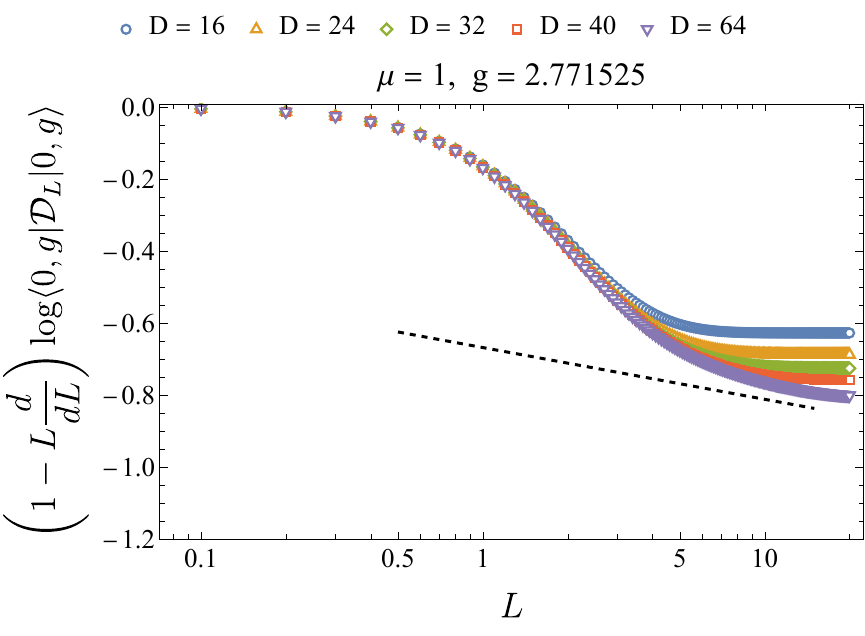}}
\subfloat[]{
	\includegraphics[width=0.48\textwidth]{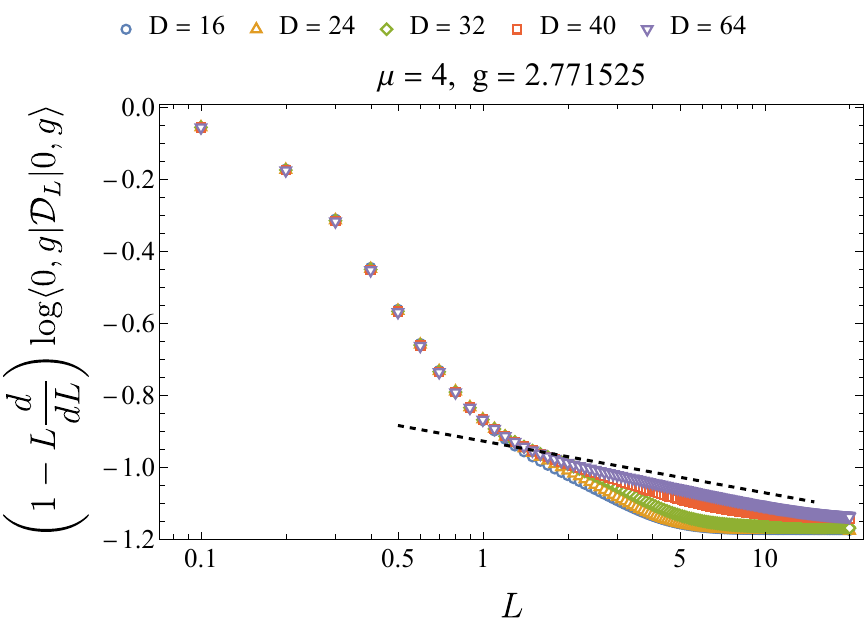}}	
	\caption{Defect vacuum expectation value from RCMPS computation as a function of the defect length $L$ at critical bulk coupling, for $\mu=1$ and $\mu=4$. The dashed line is the exact result given in eq.~\eqref{critical0pt}, shifted along the vertical axis by an (arbitrary) constant for visualization purposes.}
	\label{fig:0pt_vs_L_critical}
\end{figure}

 We first consider the expectation value of the defect. From conformal field theory, the expectation value of a defect line of length $L$ in a critical theory should scale like the two-point function of a \emph{defect creation operator}, inserted at the line defect endpoints (see \eg the appendix of ref.~\cite{Oshikawa:1996dj}). In our case, if we identify $\phi$ with the spin $\sigma$ in the 2d Ising model, the corresponding scaling exponent $\D_{+0}$ was computed in \cite{Cuomo:2024psk} and we expect
\begin{align}\label{eq:critical0pt_raw}
	\bra{0,g_c} \Dmu \ket{0,g_c} \sim e^{-E L} L^{-2\D_{+0}}\,,
\end{align}
when $L\rightarrow +\infty$ with $\D_{+0} = 1/32$ and $E$ a non-universal term. Thus should have
\begin{align}\label{critical0pt}
	\left(1-L\frac{d}{dL}\right)\log\bra{0,g_c} \Dmu \ket{0,g_c} \sim - \frac{1}{16}\log L ~ \,.
\end{align}
In Fig.~\ref{fig:0pt_vs_L_critical}, we compare this candidate scaling against RCMPS calculations, for $\mu=1$ and $\mu=4$. For sufficiently small $L$ the RCMPS results are accurate and well converged, but convergence becomes dramatically slower at large distance as expected. Even pushing the bond dimension to $D=64$, we do not see the expected large distance scaling very clearly. 

It is possible that our bond dimension is just too low, or that we would need a refined entanglement scaling analysis. It is also possible that our scaling guess \eqref{eq:critical0pt_raw} is incorrect because of additional non-universal UV effects, or substantially corrected by less relevant operators in the IR which are not decayed enough for the lengths $L$ we can reliably probe. 

\begin{figure}[htp]
	\centering
	\subfloat[]{
		\includegraphics[width=0.48\textwidth]{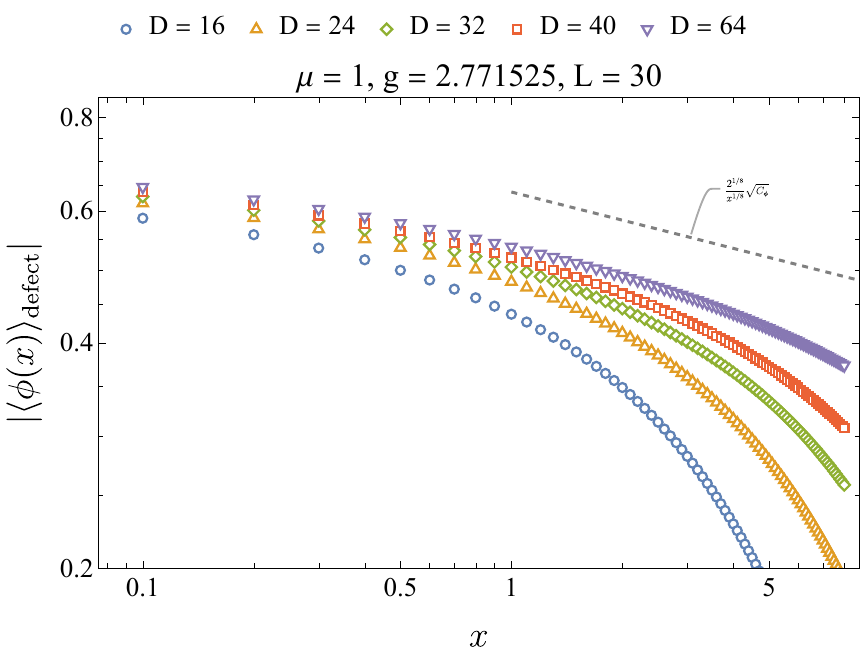}
	}
	\subfloat[]{
		\includegraphics[width=0.48\textwidth]{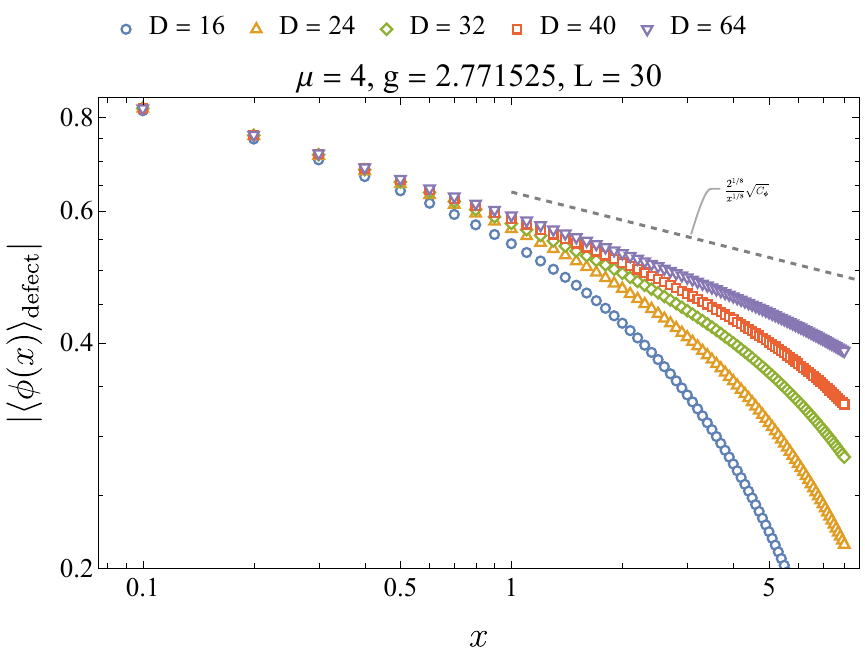}
	}	
	\caption{One-point functions obtained from RCMPS computation as a function of the distance from the defect at critical coupling, for $\mu=1$ and $\mu=4$. The dashed line is the expected exact scaling given in eq.~\eqref{critical1pt}.}
	\label{fig:1pt_vs_tau_critical}
\end{figure}

In Fig.~\ref{fig:1pt_vs_tau_critical} we show results for $|\langle \phi(x)\rangle_{\text{defect}}|$ with $\mu=1, 4$ and $L=30$. For $1\ll x \ll L$ we should again be sensitive to the critical (scaling) behavior of the model. The scaling behavior of the one-point function was computed in \cite{Cuomo:2021kfm} using Boundary Conformal Field Theory techniques (see \cite{DiFrancesco:1997nk,Cardy:2004hm} for a review): 
	\begin{align}\label{critical1pt}
		\langle \phi(x) \rangle_\text{defect}= \frac{2^{1/8}\sqrt{C_\phi}}{|x|^{1/8}}\,,
	\end{align}
where $C_\phi$ is a non-universal normalization constant which is obtained from the two-point function of the field (without defect) at large distance
\begin{equation}\label{eq:1pt_scaling_prediction}
    \bra{0,g_c} \hphi(x_1) \hphi(x_2) \ket{0,g_c} \sim \frac{C_\phi}{|x_1-x_2|^{1/4}} \,.
\end{equation}
We estimate it crudely in appendix \ref{sec:nodef2pt} by fitting the asymptotic behavior of the two-point function without the defect and get  $C_\phi\simeq 0.34$. In Fig. \ref{fig:1pt_vs_tau_critical} we observe a slightly better qualitative agreement with the scaling prediction \eqref{eq:1pt_scaling_prediction} than in the pure defect case considered before. However, without a careful finite entanglement scaling, our bond dimension remains too low to reliably fit an exponent.

\subsection{\texorpdfstring{Symmetry broken phase, coupling $g=4$}{Symmetry broken phase, coupling g=4}}
We now go deep in the symmetry broken phase. There are now two options for symmetry breaking, that correspond to very different physical situations:
\begin{itemize}
    \item The bulk can be symmetry broken in the direction favored by the defect, $ \text{sign}(\bra{0,g} \hat{\phi}(x)\ket{0,g})= - \text{sign} (\mu) $. This is what would happen spontaneously in a statistical mechanical system with a defect, initialized in the symmetric phase but then slowly cooled down. The bulk would then pick the average field value favored by the defect.
    \item The bulk can be symmetry broken ``against'' the defect, \ie $ \text{sign}(\bra{0,g} \hat{\phi}(x)\ket{0,g})= + \text{sign} (\mu)$. Physically, this would occur if symmetry breaking happens first, yielding positive $\bra{0,g} \hat{\phi}(x)\ket{0,g}$, and only then the defect coupling $\mu$ is progressively sent to positive values. The bulk field would then remain stuck ``against'' the defect.
\end{itemize}

\begin{figure}[htp]
\subfloat[]{
	\includegraphics[width=0.48\textwidth]{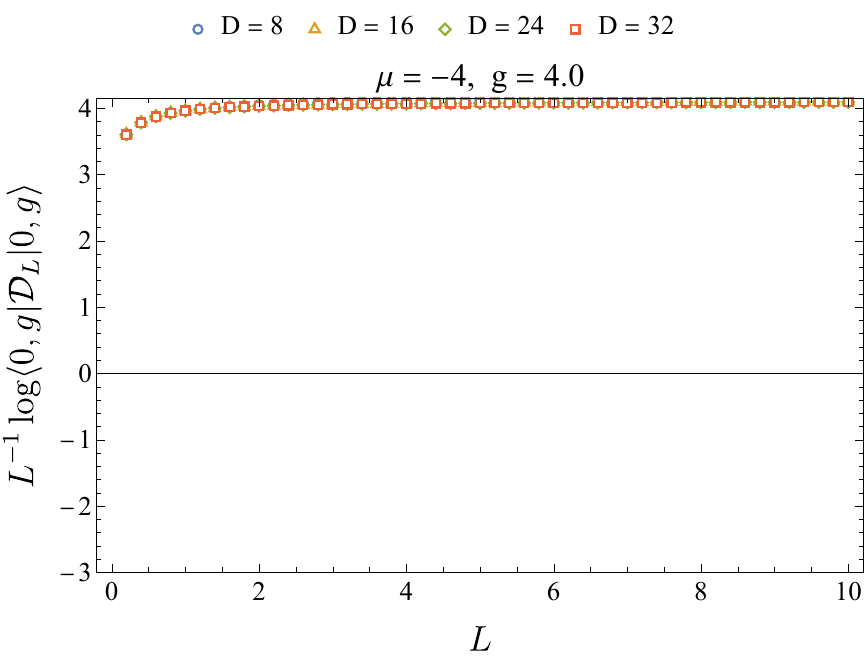}
}
\subfloat[]{
	\includegraphics[width=0.48\textwidth]{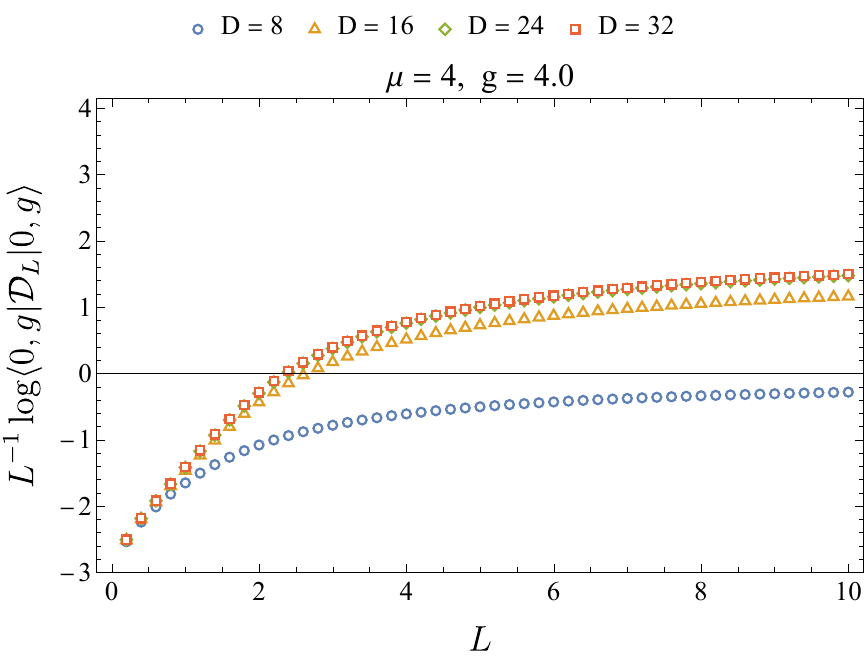}
}
	\caption{Defect vacuum expectation value from RCMPS computation as a function of the defect length $L$ in the symmetry-broken regime. In (a) symmetry breaking along the defect, in (b) symmetry breaking against the defect. In both plots $\bra{Q,R}\hphi(x)\ket{Q,R} > 0$.}
	\label{fig:0pt_vs_L_supcritical}
\end{figure}
In Fig.~\ref{fig:0pt_vs_L_supcritical}, we show the RCMPS predictions for $L^{-1} \log{\langle\mathcal{D}_L\rangle}$ as a function of $L$ in these two cases. When the symmetry is broken along the defect, the results are fairly expected: the defect expectation value increases more or less exponentially with system size, \ie $L^{-1} \log{\langle\mathcal{D}_L\rangle}$ is almost constant and positive. RCMPS results are well converged already for very small bond dimensions. 

However, when the defect is against the bulk field, the defect expectation value first decreases as a function of $L$ before increasing again, with an asymptotic exponential growth at large $L$. Convergence as a function of the bond dimension is slower, as expected from a variational method: we probe field values that are very atypical compared to those in the (approximate) ground state. Nonetheless, for $D\leq 24$ the results appear well converged.

This subtle behavior as a function of $L$ in the second case is physically interesting, and can be understood with a bit of statistical field theory intuition. When the defect is small, every point inside the defect has a field environment dominated by the bulk. This is very unfavorable, because the defect penalizes precisely the field sign favored in the bulk. However, when the defect is sufficiently large, most points inside the defect have a field environment dominated by the defect itself, with the right sign, and thus contribute positively to the defect expectation value. Intuitively, this change of behavior can happen only if $\mu$ is large enough, otherwise the pull of the bulk always dominates.

\begin{figure}[htp]
	\centering
	\includegraphics[width=0.48\textwidth]{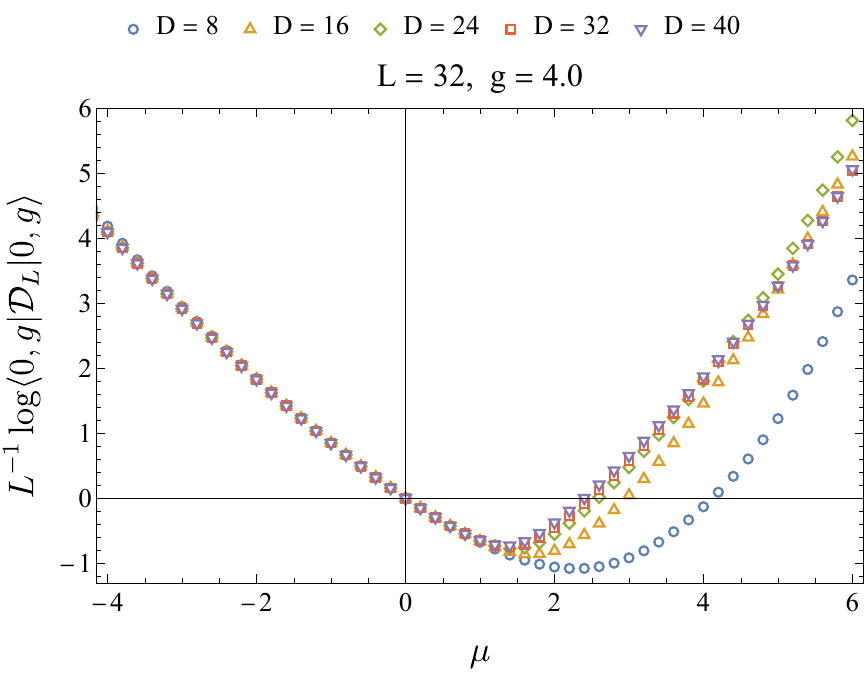}
	\caption{Asymptotic growth rate of the defect expectation value as a function of $\mu$. The results are shown for RCMPS with positive field expectation value $\bra{Q,R}\hphi(x)\ket{Q,R} > 0$.}
	\label{fig:0pt_supercritical_mu}
\end{figure}

To support this intuition, we consider the asymptotic growth rate of the defect expectation value:
\begin{equation}
    \lim_{L\rightarrow \infty} \frac{\log\bra{0,g}\Dmu \ket{0,g}}{L}\,,
\end{equation}
in the symmetry broken phase, as a function of $\mu$. Numerically, we approximate this $L \rightarrow \infty$ limit by $L=32$, which is much larger than the inverse mass scale. The corresponding results are shown in Fig. \ref{fig:0pt_supercritical_mu}.

We observe that the asymptotic growth rate of the defect expectation value is an approximate parabola that is not centered on zero but on positive $\mu$, as a clear sign of symmetry breaking. It is positive for all negative $\mu$ (this corresponds to symmetry breaking ``along'' the defect) as expected, negative for small positive $\mu$ (symmetry breaking ``against'' the defect, and a defect environment dominated by the bulk influence), and finally positive when $\mu$ is large enough (symmetry breaking ``against'' the defect, and a defect environment dominated by the defect influence).

\begin{figure}[htp]
	\centering
	\subfloat[]{
	\includegraphics[width=0.48\textwidth]{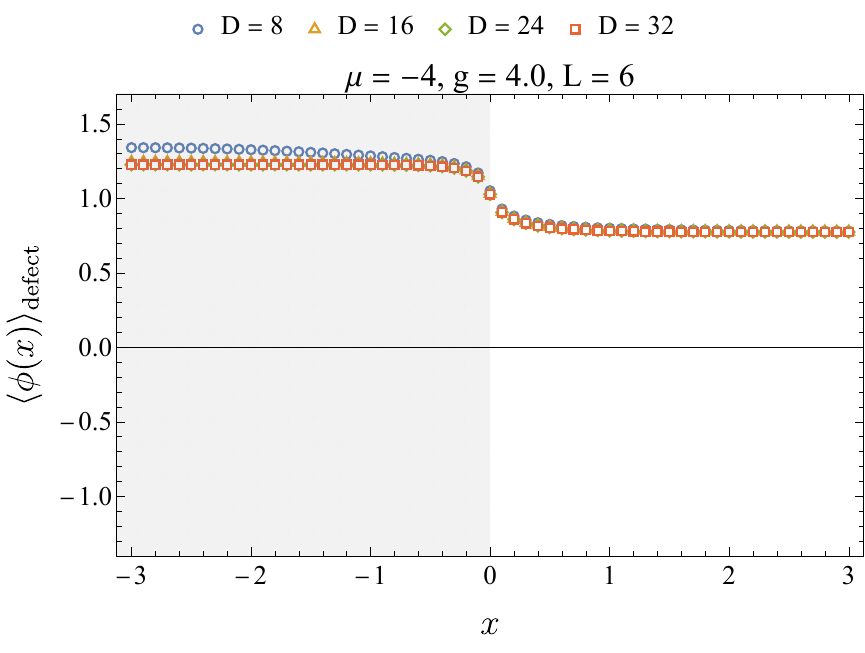}
	}
	\subfloat[]{
		\includegraphics[width=0.48\textwidth]{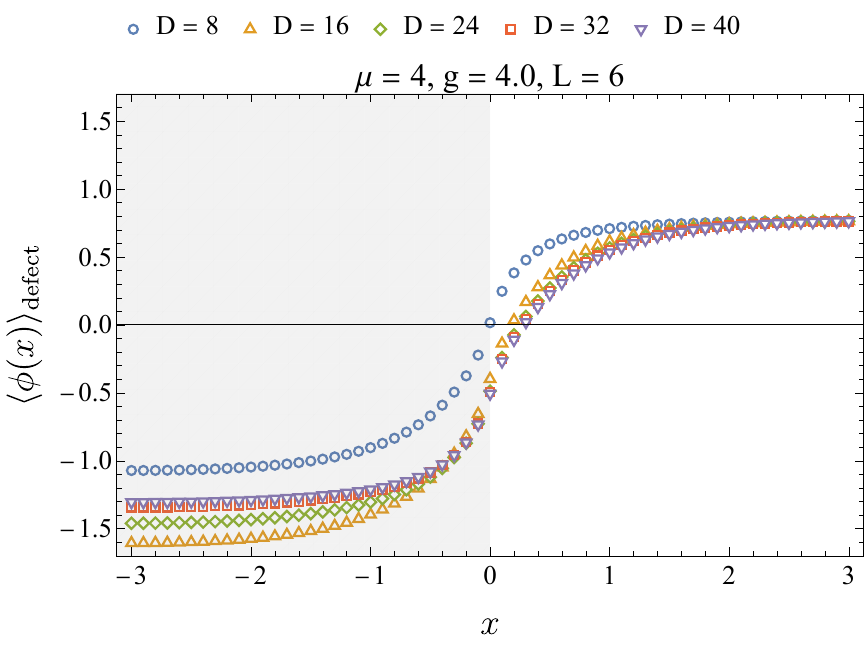}
	}
	\caption{One-point functions from RCMPS computation as a function of the distance from the defect (shaded region) in the symmetry-broken regime. In (a) symmetry breaking along the defect, in (b) symmetry breaking against the defect. In both plots $\bra{Q,R}\hphi(x)\ket{Q,R}>0$.}
	\label{fig:1pt_vs_tau_supcritical}
\end{figure}

We finally consider the $1$-point function in the presence of a defect of fixed size $L=6$ in Fig. \ref{fig:1pt_vs_tau_supcritical}, where we used again RMCPS states with $\bra{Q,R}\hphi(x)\ket{Q,R}>0$.
These results show the expected sign change in the field when moving from the defect to the bulk when the bulk is against the defect.

\section{Conclusion}
In this article, we accurately computed the expectation values in a $1+1$ dimensional scalar field theory with a magnetic line defect, using a particular variational method (RCMPS). The method is impressively efficient in many cases, albeit with a few limitations.

First, and most importantly, we could carry non-perturbative calculations without any discretizaion or needing a UV cutoff. We could also easily take the thermodynamic limit in the bulk first, for a defect of fixed size, without extra numerical cost. 

Second, we could consider a wide range of bulk couplings $g$, and go deep in the strong coupling and even symmetry-broken regime without any changes to our ansatz. We could also consider large defect coupling $\mu$, which as we recall, is impossible in perturbation theory even at fixed $g \ll 1$. Taking large defects, numerically $L$ much larger than other length-scales, brought essentially no drop in precision, and only a marginal increase in computational cost.

Putting aside the UV problems, some of the expectation values we have computed would have been difficult to estimate even on a finite lattice. Indeed, the symmetry-broken regime where the defect pulls ``against'' the bulk average, would be non-trivial to probe with the Monte-Carlo method. First, because one would have to take limits carefully to ensure the bulk remains broken against the defect during sampling. Second, because the defect expectation values are dominated by extremely rare events in that regime.

Our method shines comparatively less to estimate universal properties, where mastering the fine-grained details of the continuous model is not needed. This is particularly the case at the critical point, where our entanglement cutoff introduces an effective length-scales that breaks the scale invariance of the correlators. To estimate universal exponents, one is certainly better off with a (not necessarily fine) lattice discretization, solved with standard MPS. Universality will guarantee that the infrared physics remains identical, but the much lower computational cost of MPS would allow to reach far larger bond dimensions, and thus obtain much better precision in the IR. 

There are natural continuations of the present work. First, one could extend finite entanglement scaling techniques developed on the lattice to the continuum. This would certainly provide much better estimates of universal and non-universal quantities at criticality (even if, for universal quantities, lattice discretization will likely still dominate). Second, one could consider magnetic defects with imaginary coupling. While this presents no difficulty with RCMPS (one can take $\mu$ imaginary in all the formulas we presented), the lattice Monte-Carlo method is effectively unusable because of a growing sign problem as a function of $L$. Considering other defects (non-magnetic) like $\phi^2$ would be interesting as well, but we have not found a numerically efficient method with RCMPS yet. Finally, we could apply the present method to other bosonic field theories (like the $\phi^6$, Sine-Gordon, and Sinh-Gordon models) or even to models with multiple scalar fields (like the $O(2)$ model).

\acknowledgments{We thank G. Cuomo for discussions.
We are supported by the European Union (ERC, QFT.zip project, Grant Agreement no. 101040260). Views and opinions expressed are however those of the authors only and do not necessarily reflect those of the European Union or the European Research Council Executive Agency. Neither the European Union nor the granting authority can be held responsible for them.}

\appendix

\section{Derivation of RCMPS formulae for defects}
\label{app:tensors}

In this appendix, we present the full derivation of the RCMPS formulae given in section \ref{sec:RCMPS_defects}. Below $\ket{Q,R}$ is a RCMPS approximation to the true (interacting) ground state $\ket{0,g}$ of the $\phi^4$ model at coupling $g$.

\subsection{Expectation value of the defect operator}\label{DefectExp}

We start with the expectation value of the defect operator $\Dmu$ defined in \eqref{defectop}
\begin{align}
	\begin{split}
		\langle 0,g|\Dmu |0,g\rangle = \frac{\int   [D\phi ]\, \Dmu e^{-S_B} }{\int [D\phi ] e^{-S_B } }\,.
	\end{split}
\end{align}
Plugging in the definition of $\hphi(x)$
\begin{equation}
	\hphi(x) = \int_\mathbb{R} \upd y \, J(x-y)[\hat{a}(y) + \hat{a}^{\dagger}(y)]~, \quad J(x) = \frac{1}{2\pi} \int_\mathbb{R}\frac{\upd k}{\sqrt{2\omega_k}}e^{-i k x}\,.
\end{equation}
we can use the Baker-Campbell-Hausdorff (BCH) formula to put $\mathcal{D}_L$ in normal-ordered form
\begin{align}\label{eq:denominator}
	\begin{split}
		\Dmu   &= \; :\Dmu:\times \exp\left[\frac{\mu^{2}}{2}\int_{[-L,0]^2} \upd x \, \upd x \,^{\prime}\int_{\mathbb{R}} \upd y \; J(x-y)J(x^{\prime}-y)\right]\\
		&= \;  :\Dmu: \times \langle 0,0|\Dmu |0,0\rangle\,,
	\end{split}
\end{align}
where $\ket{0,0}$ is the ground state at coupling $0$, hence the Fock vacuum.

So far, everything was exact. Replacing $\ket{0,g}$ by its RCMPS approximation we get 
\begin{equation} 
\langle 0,g|:\Dmu :|0,g\rangle \simeq \bra{Q,R}:\Dmu : \ket{Q,R} = \mathcal{Z}_{-\mu G,-\mu G}\,,
\end{equation} and thus
\eqref{eq:denominator} gives:
\begin{equation}
    \langle 0,g|\Dmu |0,g\rangle \simeq \mathcal{Z}_{-\mu G, -\mu G}\times \exp\left[\frac{\mu^{2}}{2}\int_{\mathbb{R}} \upd y \; G(y)^2\right]\,,
\end{equation}
where $\mathcal{Z}_{j^{'},j}$ is the RCMPS generating functional defined in \eqref{RCMPS_Generator}, and $G$ is the modified ``source'', \ie 
\begin{align}
	&\mathcal{Z}_{-\mu G, -\mu G} = \tr\Big{[}\mathcal{P} \exp \int_{\mathbb{R}} \upd x \, \: \mathbb{T} - \mu G(x) (R\otimes\id + \id\otimes R^*)\Big{]}~~ \text{with}\quad G(x) := \int_{-L}^0 \upd y J(x-y)\,.
\end{align}
This path ordered exponential can be written as the solution to an ordinary differential equation (ODE). Putting the ODE in superoperator form as in the main text, we finally get
\begin{align}
	\bra{0,g}\Dmu\ket{0,g} \simeq \lim_{x \to \infty} \, \mathrm{tr}\,[\rho(x)]~, ~~ \text{for the initial condition}~~~ \qquad \lim_{x\to -\infty} \rho(x)= \rho_{0}\,.
\end{align}
and 
\begin{equation}\label{eq:denODE_bis}
	\frac{\upd}{\upd x} \rho(x) = \mathcal{L}\cdot \rho(x) - \mu G(x) \left[ R\rho(x) + \rho(x) R^\dagger \right] + \frac{\mu^{2}}{2} G^{2}(x) \rho(x)\,,
\end{equation}
where we recall that
\begin{align}
	\mathcal{L} \cdot \rho = Q\, \rho + \rho \, Q^{\dagger} + R \, \rho \, R^{\dagger}\,.
\end{align}

\subsection{One-point functions in the defect model}

\subsubsection{Vertex operators in the defect model}\label{sec:DefectVertex}

Next, we consider the vertex operator in the full defect theory \eqref{eq:Defect_vertex}, which we differentiate in \ref{app:monomial} to obtain expectation values of one point functions in the $\phi^4$ model in presence of the defect. It is defined as
\begin{align}\label{eq:Defect_vertex_bis}
    \langle V_b (x) \rangle_{\text{defect}} = \frac{\bra{0,g} :e^{b\hat{\phi}(x)}: \text{e}^{-\mu \int_{-L}^{0} \hat{\phi}}\ket{0,g}}{\bra{0,g}\text{e}^{-\mu \int_{-L}^{0} \hat{\phi}}\ket{0,g}}\,.
\end{align}
We first focus on the numerator, as the denominator has just been computed.

The first step, again exact, is to normal-order the operator being evaluated using the Baker-Campbell-Hausdorff formula:
\begin{equation}
\begin{split}
    : e^{\, b \, \hat{\phi}(x)}: e^{-\mu \int_{-L}^0 \upd y \, \: \hphi(y)} = &\,: e^{ b \, \hphi(x)-\mu \int_{-L}^0 \upd y \, \: \hphi(y) }: \; \bra{0,0}\Dmu \ket{0,0}\\
    & \times   \exp\left[- b\,\mu\,\int_{\mathbb{R}} \upd y  \:J(x-y)G(y)\right]\,,
    \end{split}
\end{equation}
where we have explicitly separated the $\bra{0,0}\Dmu \ket{0,0}$ for convenience as it also appears in the denominator and thus ultimately cancels out.

Replacing the true ground state by its variational approximation yields
\begin{equation}
\begin{split}
    \bra{0,g} :e^{b\hat{\phi}(x)}: \text{e}^{-\mu \int_{-L}^{0} \hat{\phi}}\ket{0,g} \simeq &\,\mathcal{Z}_{s_x,s_x} \; \bra{0,0}\Dmu \ket{0,0}\\
    & \times   \exp\left[- b\,\mu\,\int_{\mathbb{R}} \upd y \:J(x-y)G(y)\right]\,,
\end{split}
\end{equation}
where 
\begin{equation}{\label{eq:smearedsource}}
	s_x(y) := b\,J(x-y) - \mu\int_{-L}^0 \upd z \: J(z-y) = b\, J(x-y) - \mu \,  G(y) \,.
\end{equation}
Again, we can write $\mathcal{Z}_{s_x,s_x}$ as the solution of an ODE to get
\begin{equation}\label{eq:vertsolution}
	\bra{Q,R} : e^{\, b \, \hphi(x)}: \Dmu \ket{Q,R} = \langle 0,0|\Dmu |0,0\rangle \, \lim_{y\to +\infty}\tr[\rho(y)]\,,
\end{equation}
with
\begin{align}\label{eq:defectvertexODE}
	\frac{\upd}{\upd y} \rho(y) &= \mathcal{L}\cdot \rho(x) + s_x(y)  \left[R\rho(x) + \rho(x) R^\dagger\right]- b \, \mu \, G(y)J(x-y)\rho(y)\,,
\end{align}
and with the initial condition $\lim_{y\rightarrow-\infty}\rho(y) = \rho_{0}$ with $\tr[\rho_0]=1$. Finally, we can put numerator and denominator together, which cancels the $\bra{0,0} \Dmu\ket{0,0}$ term and we get
\begin{equation}\label{eq:vertex_as_trace}
    \langle V_b (x) \rangle_{\text{defect}} \simeq \lim_{y\rightarrow \infty}  \frac{\tr[\rho(y)]}{\tr[\tilde{\rho}(y)]}\,,
\end{equation}
where $\tilde{\rho}(y)$ is the solution of \eqref{eq:defectvertexODE} with $b=0$ (or equivalently of \eqref{eq:denODE_bis} without $G^2$ term).

\subsubsection{Field monomials in the defect model} \label{app:monomial}

With explicit expressions for defect vertex operators, we may now compute expectation values of field monomials in the full defect theory. The latter are defined as:
\begin{equation}\label{eq:npointdef}
	\langle : \phi^{n}(x): \rangle_{\text{defect}} = \frac{ \langle 0,g | : \hphi^{n}(x): e^{-\mu \int_{-L}^0 \upd x \, \: \hphi(x)} |0,g\rangle}{ \langle 0,g | e^{-\mu \int_{-L}^0 \upd x \, \: \hphi(x)}|0,g \rangle}\,.
\end{equation}
Again, we already have the denominator from \eqref{DefectExp}. For the numerator, we can differentiate the defect vertex operator w.r.t $b$,
\begin{align}\label{eq:npointdef2}
	\begin{split}
		\langle 0,g| : \hphi^{n}(x): e^{-\mu \int_{-L}^0d x \, \: \hphi(x)} |0,g\rangle  
		& = 	\frac{\partial
			^{n}}{\partial b^{n}}\langle 0,g| : e^{\, b \, \hphi(x)}: \Dmu |0,g\rangle \Big|_{b=0}\,.
	\end{split}
\end{align}
The strategy is now to forward differentiate the ordinary differential equation giving us $\rho(y)$ appearing in \eqref{eq:vertex_as_trace}
\begin{align}
	\bra{0,g} : \hphi^{n}(x): e^{-\mu \int_{-L}^0 \upd x \, \: \hphi(x)} \ket{0,g} \, &\simeq \, \bra{Q,R} :\hphi^{n}(x): e^{-\mu \int_{-L}^0 \upd x \, \: \hphi(x)} \ket{Q,R} \\
    \, &= \, \langle 0,0|\Dmu |0,0\rangle \times \, \lim_{y\to +\infty}\tr\left[\rho^{(n)}(y)\right]\,, \label{eq:monomial_from_rho_l2}
\end{align}
where $\rho^{(n)}(y):= \partial_b^n \rho(y) |_{b=0}$ and $\rho$ obeys the ODE \eqref{eq:defectvertexODE}. The matrices $\rho^{(k)}(y)$ for $0\leq k\leq n$ obey the triangular system of matrix ODEs
\begin{align}\label{eq:numODE}
	\frac{\upd}{\upd y} \rho^{(k)}(y) &= \mathcal{L} \cdot \rho^{(k)}(y) + k J(x-y)\left[R\rho^{(k-1)}(y) + \rho^{(k-1)}(y) R^\dagger\right] \\
    &- \mu G(y)\left[R\rho^{(k)}(y) + \rho^{(k)}(y) R^\dagger\right] - k\, \mu \, J(x-y)G(y)\rho^{(k-1)}(y)\,,
\end{align}
with the initial conditions, $ \lim_{y\rightarrow-\infty}\rho^{(0)}(y) = \rho_{0}$ and $\lim_{y\rightarrow-\infty}\rho^{(n)}(y) = 0$ for $k > 0$. Here, again the factor $\langle 0,0|\Dmu |0,0\rangle$ arises both in the numerator \eqref{eq:monomial_from_rho_l2} and the denominator \eqref{eq:denominator} and hence drops out in the computation of monomials \eqref{eq:npointdef}. Putting all together, we find
\begin{equation}\label{eq:monomial_final_as_trace}
	\langle : \phi^{n}(x): \rangle_{\text{defect}} = \lim_{y\to +\infty} \frac{\tr[\rho^{(n)}(y)]}{\tr[\rho^{(0)}(y)]}\,.
\end{equation}

\subsubsection{Normalization tricks} \label{app:normalization_tricks}
The vertex operator and field monomial expectation values in the presence of the defect given in \eqref{eq:vertex_as_trace} and \eqref{eq:monomial_final_as_trace} are ratios of traces, each of which is typically growing exponentially with the size $L$ of the defect. Thus, computing each term independently and taking the ratio quickly gives large numerical errors in double precision, especially when numerically taking the semi-infinite defect line limit. This should be avoided.

A simple way to make things better behaved is to periodically normalize both the matrix in the numerator and the matrix in the denominator by the trace of the denominator as the ODE is being solved. For example, in the case of the computation of the field expectation
\begin{equation}\label{eq:field_as_trace}
	\langle : \phi(x): \rangle_{\text{defect}} = \lim_{y\to +\infty} \frac{\tr[\rho^{(1)}(y)]}{\tr[\rho^{(0)}(y)]}\,,
\end{equation}
we can pick a number of checkpoints $y_k$ (either a priori, or each time $\tr\left[\rho^{(0)}(y)\right]$ gets too large or too small) and renormalize both $\rho^{(0)}(y_k)$ and $\rho^{(1)}(y_k)$:
\begin{align}
    \rho^{(0)}(y_k^+) &= \frac{\rho^{(0)}(y_k)}{\tr\left[\rho^{(0)}(y_k)\right]}\,, \\
    \rho^{(1)}(y_k^+) &= \frac{\rho^{(1)}(y_k)}{\tr\left[\rho^{(0)}(y_k)\right]} \,.
\end{align}
Inserting these normalizations do not change the value of the expectation value \eqref{eq:field_as_trace}.

Alternatively, one could define the continuously normalized matrices
\begin{align}
    \sigma^{(0)}(y) &= \frac{\rho^{(0)}(y)}{\tr\left[\rho^{(0)}(y)\right]} \\
    \sigma^{(1)}(y) &= \frac{\rho^{(1)}(y)}{\tr\left[\rho^{(0)}(y)\right]} \,,
\end{align}
which indeed obey a system of closed ODEs. However these ODEs are non-linear, and we found it numerically preferable to use the previous periodic normalization strategy. No matter which strategy is used, one can compute expectation values for large or even numerically semi-infinite defects without suffering from instabilities of precision loss.

\subsection{Strategy for more general expectation values} \label{app:RCMPS_2point}
Following the same techniques, one can compute expectation values of arbitrary strings of vertex operators or normal-ordered monomials in the presence of the defect. The steps are the same as before, and we only outline the general strategy here. One first considers expectation values of products of vertex operators
\begin{align}\label{eq:Defect_vertex_multiple}
    \langle V_{b_1} (x_1) V_{b_2}(x_2) \dots V_{b_n}(x_n) \rangle_{\text{defect}} = \frac{\bra{0,g} :e^{b_1\hat{\phi}(x_1)}\!: \, :e^{b_2\hat{\phi}(x_2)}\!: \, \dots \, :e^{b_n\hat{\phi}(x_n)}\! : \; \text{e}^{-\mu \int_{-L}^{0} \hat{\phi}}\ket{0,g}}{\bra{0,g}\text{e}^{-\mu \int_{-L}^{0} \hat{\phi}}\ket{0,g}}~ \,\,,
\end{align}
for $x_1  <x_2 <\cdots < x_n$. Then, focusing on the numerator, we use the Baker-Campbell-Hausdorff formula iteratively to put all the fields under the same exponential
\begin{equation}
\begin{split}
    :e^{b_1\hat{\phi}(x_1)}\!: \, :e^{b_2\hat{\phi}(x_2)}\!: \, \dots \, :e^{b_n\hat{\phi}(x_n)}\! : \text{e}^{-\mu \int_{-L}^{0}\hat{\phi}} &=\\ \mathcal{N}(b_1,b_2,\dots,b_n,x_1,x_2,\cdots,x_n) \; \times \; &:e^{b_1\hat{\phi}(x_1) + b_2\hat{\phi}(x_2) + \dots + b_n\hat{\phi}(x_n) -\mu \int_{-L}^{0}\hat{\phi}} : \,,
\end{split}
\end{equation}
where $\mathcal{N}$ is an explicit scalar contribution coming from the normal ordering. The expectation value of the remaining normal-ordered exponential on a RCMPS is then just $\mathcal{Z}_{s,s}$ with
\begin{equation}
    s(y) := \sum_{j=1}^n b_j\, J(x_j-y) - \mu \,  G(y) \,,
\end{equation}
and can thus be obtained by solving a simple linear ODE like before. 

Then, to compute expectation values of normal-ordered field monomials at different points, one may just differentiate with respect to the $b$'s
\begin{equation}
    \langle :{\phi}^{k_1}(x_1): \, :{\phi}^{k_2}(x_2): \, \dots \, :{\phi}^{k_n}(x_n): \rangle_{\text{defect}}\, =\partial_{b_1}^{k_1} \partial_{b_2}^{k_2} \dots \partial_{b_n}^{k_n}  \langle V_{b_1} (x_1) V_{b_2}(x_2) \dots V_{b_n}(x_n) \rangle_{\text{defect}} \bigg|_{b=0} \,. 
\end{equation}

As an illustration, following these steps one gets:
\begin{equation}
    \langle\phi(x_1)\phi(x_2)\rangle_\text{defect} = \lim_{y\rightarrow + \infty}\frac{\tr[\rho^{(1,1)}(y)]}{\tr[\rho^{(0,0)}(y)]}\,,
\end{equation}
with
\begin{align}
	\begin{split}
		\frac{\upd \rho^{(0,0)}(y)}{\upd y} &= [\mathcal{L}- \mu G(y)\mathcal{R}]\cdot \rho^{(0,0)}(y)\\
		\frac{\upd \rho^{(1,0)}(y)}{\upd y}&= [\mathcal{L}- \mu G(y)\mathcal{R}]\cdot \rho^{(1,0)}(y) + [J(x_1-y)(\mathcal{R}-\mu \, G(y)\id)]\cdot \rho^{(0,0)}(y)\\
		\frac{\upd \rho^{(0,1)}(y)}{\partial y}&= [\mathcal{L}- \mu \, G(y)\mathcal{R}]\cdot \rho^{(0,1)}(y) + [J(x_2-y)(\mathcal{R}-\mu G(y)\id)]\cdot \rho^{(0,0)}(y)\\
		\frac{\upd \rho^{(1,1)}(y)}{\upd y} &= [\mathcal{L}- \mu \, G(y)\mathcal{R}]\cdot \rho^{(1,1)}(y) + [J(x_2-y)(\mathcal{R}-\mu \, G(y)\id)]\cdot \rho^{(1,0)}(y)\\
		& \quad + [J(x_1-y)(\mathcal{R}-\mu \, G(y)\id)]\cdot \rho^{(0,1)}(y) + J(x_1-y)J(x_2-y)\rho^{(0,0)}(y)\,,
	\end{split}
\end{align}
with $\mathcal{R}\cdot \rho = R \rho + \rho R^\dagger$ and the initial conditions $\lim_{y\to -\infty} \rho^{(0,0)}(y)= \rho_{0}$ and $\lim_{y\to -\infty} \rho^{(i,j)}(y)= 0$. 

While we do not show the result of two-point functions with defects in the main text, we verified numerically that they converged approximately as fast as the defect one-point functions as a function of $D$ and matched perturbation theory at sufficiently small coupling.

\section{No-defect two-points correlators}\label{sec:nodef2pt}
In this appendix we consider two-point function of $\phi$ in the no-defect theory
\begin{align}
	\langle \phi(x_1) \phi(x_2)\rangle_\text{bulk}:= \bra{0,g} \hat{\phi}(x_1)\hphi(x_2)\ket{0,g}\,.
\end{align}

\subsection{Perturbative regime}

\begin{figure}[htp]
	\centering
	\subfloat[]{
		\includegraphics[width=0.48\textwidth]{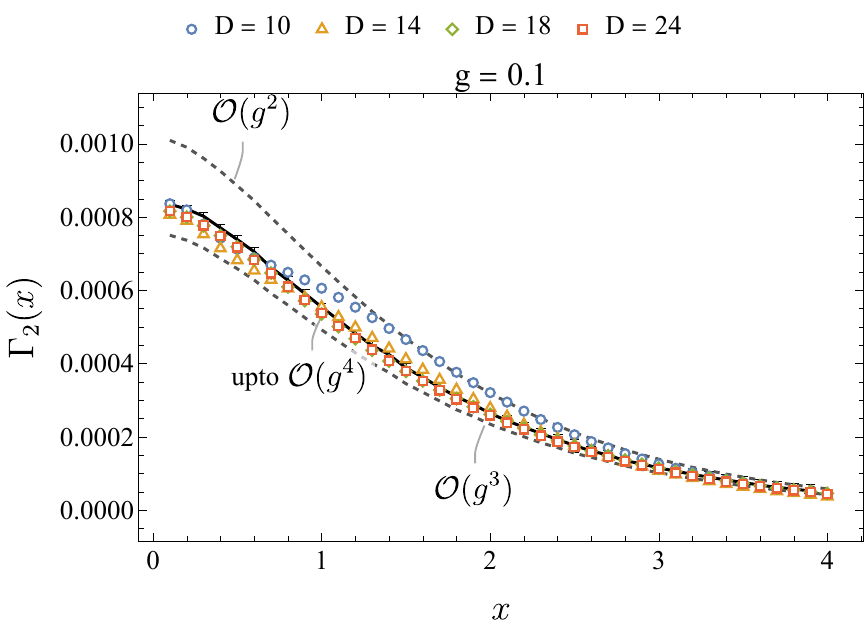}}
		\subfloat[]{
			\includegraphics[width=0.48\textwidth]{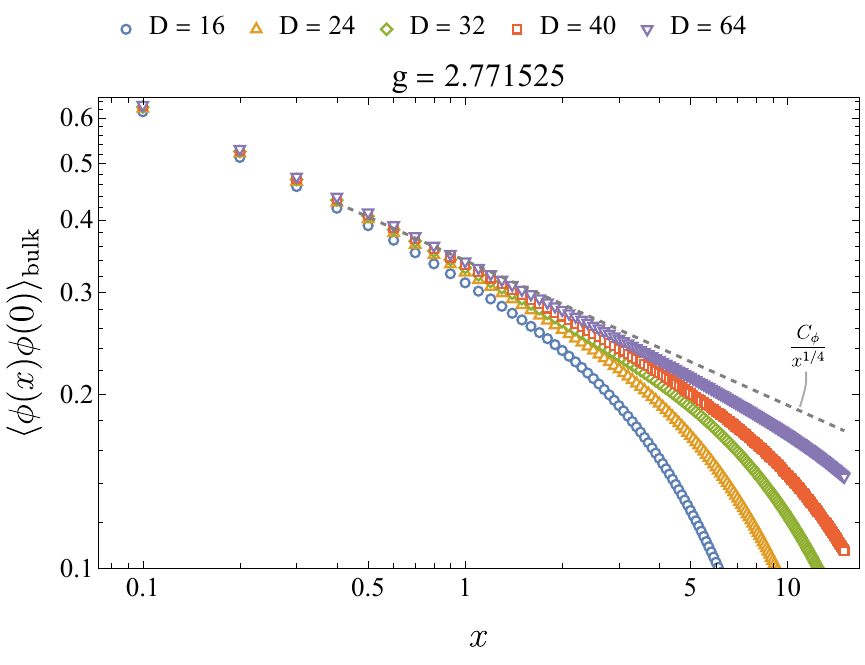}}
	\caption{In (a), up to $\order(g^4)$ perturbative prediction for $ \Gamma_2(x) \equiv \bra{0,0} \phi(0) \phi(x)\ket{0,g}-\bra{0,g} \phi(0) \phi(x)\ket{0,0}$ vs. RCMPS (colored markers). Monte Carlo error bars are smaller than the size of the data points. In (b), no-defect two-point functions from RCMPS computation as a function of the distance from the defect at critical coupling. The dashed line is eq.~\eqref{twopt_nodef_critical} with $C_\phi \simeq  0.34$.}
	\label{fig:2pt_vstau_nodef}
\end{figure}

In perturbation theory, when $\mu=0$, this two-point function has the following Feynman diagram expansion (see \cite{Serone:2018gjo,Serone:2019szm} for the higher-order computation)
\begin{align}\label{2pt_phi_pert_nodef}
	\bra{0,g} &\phi(x_1) \phi(x_2)\ket{0,g}=\nonumber\\
	&\begin{tikzpicture}[baseline,valign]
		\draw[thick] (0, 0) -- (1.4, 0);
	\end{tikzpicture}
	+\tfrac{1}{6}\begin{tikzpicture}[baseline,valign]
		\draw[thick] (0, 0) -- (1.4, 0);
		\draw[thick] (0.2, 0.0) to[out=90,in=90] (1.2, 0.0);
		\draw[thick] (0.2, 0.) to[out=-90,in=-90] (1.2, 0.0);
		\node at (0.2, 0.) [rcirc] {};
		\node at (1.2, 0.) [rcirc] {};
	\end{tikzpicture}
	+\tfrac{1}{12}\begin{tikzpicture}[baseline,valign]
		\draw[thick] (0.2, 0.6) -- (1.2, 0.6);
		\draw[thick] (0., 0.0) -- (1.4, 0.);
		\draw[thick] (0.2, 0.6) to[out=30,in=150] (1.2, 0.6);
		\draw[thick] (0.2, 0.6) to[out=-30,in=-150] (1.2, 0.6);
		\draw[thick] (0.2, 0.6) to[out=-90,in=-180] (0.7, 0.0);
		\draw[thick] (1.2, 0.6) to[out=-90,in=0] (0.7, 0.0);
		\node at (0.2, 0.6) [rcirc] {};
		\node at (1.2, 0.6) [rcirc] {};
		\node at (0.7, 0.0) [rcirc] {};
	\end{tikzpicture}
	+\tfrac{1}{4}\begin{tikzpicture}[baseline,valign]
		\draw[thick] (0., 0.0) -- (1.4, 0.);
		\draw[thick] (0.2, 0.0) to[out=90,in=180] (0.7, 0.6);
		\draw[thick] (0.7, 0.6) to[out=0,in=90] (1.2, 0.0);
		\draw[thick] (0.2, 0.) to[out=0,in=-90] (.7, 0.6);
		\draw[thick] (0.7, 0.6) to[out=-90,in=180] (1.2, 0.0);
		\node at (0.2, 0.0) [rcirc] {};
		\node at (1.2, 0.0) [rcirc] {};
		\node at (0.7, 0.6) [rcirc] {};
	\end{tikzpicture}+\tfrac{1}{24}\begin{tikzpicture}[baseline,valign]
		\draw[thick] (0.2, 0.6) -- (1.2, 0.6);
		\draw[thick] (0., -.4) -- (1.4, -.4);
		\draw[thick] (0.2, 0.6) to[out=30,in=150] (1.2, 0.6);
		\draw[thick] (0.2, 0.6) to[out=-30,in=-150] (1.2, 0.6);
		\draw[thick] (0.2, 0.6) to[out=-90,in=-180] (0.7, 0.0);
		\draw[thick] (1.2, 0.6) to[out=-90,in=0] (0.7, 0.0);
		\node at (0.2, 0.6) [rcirc] {};
		\node at (1.2, 0.6) [rcirc] {};
		\node at (0.7, 0.0) [rcirc] {};
		\node at (0.7, -.4) [rcirc] {};
		\draw[thick] (0.7, 0.) to[out=0,in=0] (.7, -0.4);
		\draw[thick] (0.7, 0.) to[out=180,in=180] (.7, -0.4);
	\end{tikzpicture}
	+\tfrac{1}{12}\begin{tikzpicture}[baseline,valign]
		\draw[thick] (0.2, 0) -- (1.2, 0);
		\draw[thick] (0.2, 0.0) to[out=30,in=150] (1.2, 0.0);
		\draw[thick] (0.2, 0.) to[out=-30,in=-150] (1.2, 0.0);
		\draw[thick] (0.2, -0.7) to[out=90,in=90] (1.2, -0.7);
		\draw[thick] (0.2, -0.7) to[out=-90,in=-90] (1.2, -0.7);
		\draw[thick] (0, -0.7) -- (.2, -0.7);
		\draw[thick] (1.2, -0.7) -- (1.4, -0.7);
		\draw[thick] (1.2, 0.) -- (1.2, -0.7);
		\draw[thick] (.2, 0.) -- (.2, -0.7);
		\node at (0.2, 0.) [rcirc] {};
		\node at (1.2, 0.) [rcirc] {};
		\node at (1.2, -0.7) [rcirc] {};
		\node at (0.2, -0.7) [rcirc] {};
	\end{tikzpicture}\nn \\
	&
	+\tfrac{1}{36}\begin{tikzpicture}[baseline,valign]
		\draw[thick] (0, 0) -- (2.6, 0);
		\draw[thick] (0.2, 0.0) to[out=90,in=90] (1.2, 0.0);
		\draw[thick] (1.4, 0.) to[out=-90,in=-90] (2.4, 0.0);
		\draw[thick] (1.4, 0.0) to[out=90,in=90] (2.4, 0.0);
		\draw[thick] (0.2, 0.) to[out=-90,in=-90] (1.2, 0.0);
		\node at (0.2, 0.) [rcirc] {};
		\node at (1.2, 0.) [rcirc] {};
		\node at (1.4, 0.) [rcirc] {};
		\node at (2.4, 0.) [rcirc] {};
	\end{tikzpicture}
	+\tfrac{1}{8}\begin{tikzpicture}[baseline,valign]
		\draw[thick] (0., 0.0) -- (1.4, 0.);
		\draw[thick] (0.2, 0.0) to[out=90,in=180] (0.7, 0.6);
		\draw[thick] (0.7, 0.6) to[out=0,in=90] (1.2, 0.0);
		\draw[thick] (0.2, 0.) to[out=0,in=-90] (.7, 0.6);
		\draw[thick] (0.7, 0.6) to[out=-90,in=180] (1.2, 0.0);
		\draw[thick] (0.2, 0.0) to[out=270,in=90] (0.7, -0.6);
		\draw[thick] (1.2, 0.0) to[out=270,in=90] (0.7, -0.6);
		\draw[thick] (0., -.6) -- (1.4, -0.6);
		\node at (0.2, 0.0) [rcirc] {};
		\node at (1.2, 0.0) [rcirc] {};
		\node at (0.7, 0.6) [rcirc] {};
		\node at (0.7, -0.6) [rcirc] {};
	\end{tikzpicture}
	+\tfrac{1}{4}\begin{tikzpicture}[baseline,valign]
		\draw[thick] (0., 0.0) -- (0.2, 0.);
		\draw[thick] (1.2, 0.0) -- (1.4, 0.);
		\draw[thick] (0.2, 0.0) -- (.7, 0.5) -- (1.2, 0.) -- (.7, -0.5) -- (.2, 0.0);
		\draw[thick] (0.7, 0.5) -- (0.7, -0.5);
		\draw[thick] (0.2, 0.0) to[out=0,in=90] (0.7, -0.5);
		\draw[thick] (1.2, 0.0) to[out=180,in=-90] (0.7, 0.5);
		\node at (0.2, 0.0) [rcirc] {};
		\node at (1.2, 0.0) [rcirc] {};
		\node at (0.7, 0.5) [rcirc] {};
		\node at (0.7, -0.5) [rcirc] {};
	\end{tikzpicture}
	+\tfrac{1}{8}\begin{tikzpicture}[baseline,valign]
		\draw[thick] (0.2, 0.0) to[out=30,in=150] (1.2, 0.0);
		\draw[thick] (0.2, 0.0) to[out=-30,in=-150] (1.2, 0.0);
		\draw[thick] (0.2, 0.0) to[out=-150,in=150] (0.2, -0.7);
		\draw[thick] (1.2, 0.0) to[out=-30,in=30] (1.2, -0.7);
		\draw[thick] (0, -0.7) -- (1.4, -0.7);
		\draw[thick] (1.2, -0.7) -- (1.4, -0.7);
		\draw[thick] (1.2, 0.) -- (1.2, -0.7);
		\draw[thick] (.2, 0.) -- (.2, -0.7);
		\node at (0.2, 0.) [rcirc] {};
		\node at (1.2, 0.) [rcirc] {};
		\node at (1.2, -0.7) [rcirc] {};
		\node at (0.2, -0.7) [rcirc] {};
	\end{tikzpicture}
	+\tfrac{1}{4}\begin{tikzpicture}[baseline,valign]
		\draw[thick] (0.2, 0.0) to[out=30,in=150] (1.2, 0.0);
		\draw[thick] (0.2, 0.0) -- (1.2, 0.0);
		\draw[thick] (0.2, 0.0) -- (1.2, -0.7);
		\draw[thick] (1.2, 0.0) -- (0.2, -0.7);
		\draw[thick] (0, -0.7) -- (1.4, -0.7);
		\draw[thick] (1.2, -0.7) -- (1.4, -0.7);
		\draw[thick] (1.2, 0.) -- (1.2, -0.7);
		\draw[thick] (.2, 0.) -- (.2, -0.7);
		\node at (0.2, 0.) [rcirc] {};
		\node at (1.2, 0.) [rcirc] {};
		\node at (1.2, -0.7) [rcirc] {};
		\node at (0.2, -0.7) [rcirc] {};
	\end{tikzpicture}\nonumber\\
	&+ \order(g^5)\,.
\end{align}
The Feynman diagrams in this expression are evaluated numerically, and the result is compared with RCMPS predictions in fig.~\ref{fig:2pt_vstau_nodef}, which shows good agreement between the two methods.

\subsection{Critical coupling regime}
We now tune the bulk coupling to its critical value, $g_c\simeq 2.771525$ \cite{Delcamp:2020hzo,Tilloy:2021yre,Tilloy:2021hhb,Tilloy:2022kcn}. In fig.~\ref{fig:2pt_vstau_nodef} we show results for $\bra{0,g_c}\hphi(x)\hphi(0)\ket{0,g_c}$. In the scaling region the result is compatible with the power-law decay for critical Ising model \cite{DiFrancesco:1997nk}, \ie
	\begin{align}\label{twopt_nodef_critical}
	\langle \phi(x) \phi(0)\rangle_\text{bulk} &= \frac{C_\phi}{|x|^{1/4}}\,,
\end{align}
where $C_\phi \simeq 0.34$ is a non-universal coefficient which we evaluate numerically from the scaling region of the plot in Fig.~\ref{fig:2pt_vstau_nodef}.

\bibliographystyle{apsrev4-1}
\bibliography{main}

\end{document}